\documentclass[longauth]{aa}
\usepackage{graphicx}
\usepackage{txfonts}
\usepackage[colorlinks=true,citecolor=blue,backref=page]{hyperref}
\usepackage{xspace}
\usepackage{supertabular}
\usepackage{longtable}

\newcommand{\mps}{$\mathrm{m\,s^{-1}}$}

\newcommand{\logrhk}{$\log R'_{HK}$}
\newcommand{\host}{HD 20329}
\newcommand{\tess}{\textit{TESS}\xspace}
\newcommand{\pytransit}{\texttt{PyTransit}\xspace}
\newcommand{\ldtk}{\texttt{LDTK}\xspace}
\newcommand{\celerite}{\texttt{Celerite}\xspace}

\begin{document} 

   \title{HD 20329b: An ultra-short-period planet around a solar-type star found by \textit{TESS}}

\author{F. Murgas\inst{ \ref{iac},\ref{ull}}
          \and
          G. Nowak\inst{ \ref{iac},\ref{ull}}
          \and
          T. Masseron\inst{ \ref{iac},\ref{ull}}
          \and
          H. Parviainen\inst{ \ref{iac},\ref{ull}}
          \and
          R. Luque\inst{\ref{iaa}, \ref{uchicago}}
          \and
          E. Pall\'{e}\inst{ \ref{iac},\ref{ull}}
          \and
          Judith Korth\inst{\ref{chalmers1}}
          \and
          I. Carleo\inst{ \ref{iac},\ref{ull}}
          \and
          Sz. Csizmadia\inst{\ref{dlr}}
          \and
          E.~Esparza-Borges\inst{ \ref{iac},\ref{ull}}
          \and
          Ahlam Alqasim\inst{\ref{ucl}}
          \and
          William D. Cochran\inst{ \ref{mcdonobs}, \ref{utexas}}
          \and
          Fei Dai\inst{\ref{pasadena}}
          \and
          Hans J. Deeg\inst{ \ref{iac},\ref{ull}}
          \and
          D. Gandolfi\inst{\ref{utorino}}
          \and
          Elisa Goffo\inst{\ref{utorino}, \ref{tautenburg}}
          \and
          Petr Kab\'{a}th\inst{\ref{czechacad}}
          \and
          K. W. F. Lam\inst{\ref{dlr}}
          \and
          John Livingston\inst{\ref{astrobiosawa}, \ref{noao}, \ref{sokendai}}
          \and
          Alexandra Muresan\inst{ \ref{chalmers1}}
          \and
          H. L. M. Osborne\inst{\ref{ucl}}
          \and
          Carina M. Persson\inst{ \ref{chalmers2}}
          \and
          L. M. Serrano\inst{\ref{utorino}}
          \and
          Alexis M. S. Smith\inst{\ref{dlr}}
          \and
          Vincent Van Eylen\inst{\ref{ucl}}
          \and
          J.~Orell-Miquel\inst{ \ref{iac},\ref{ull}}
          \and
          Natalie R. Hinkel\inst{\ref{southwestRI}}
          \and
          D. Gal\'{a}n\inst{\ref{ull}}
          \and
          M. Puig-Subir\`{a} \inst{ \ref{iac},\ref{ull}}
          \and
          M. Stangret\inst{ \ref{iac},\ref{ull}, \ref{inaf}}
          \and
          A. Fukui\inst{ \ref{iac},\ref{utokyo1}}
          \and
          T. Kagetani\inst{\ref{utokyo2}}
          \and
          N. Narita\inst{ \ref{utokyo1},\ref{iac}, \ref{astrobiosawa}}
          \and
          David R. Ciardi\inst{\ref{nasaipac}}
          \and
          Andrew W. Boyle\inst{\ref{nasaipac}}
          \and
          Carl Ziegler\inst{\ref{austin}}
          \and
          C\'{e}sar Brice\~{n}o\inst{\ref{tololo}}
          \and
          Nicholas Law\inst{\ref{unc}}
          \and
          Andrew W. Mann\inst{\ref{unc}}
          \and
          Jon M. Jenkins\inst{\ref{nasaames}}
          \and
          David W. Latham\inst{\ref{harvardsmith}}
          \and
          Samuel N. Quinn\inst{\ref{harvardsmith}}
          \and
          G. Ricker\inst{\ref{kavlimit}}
          \and
          S. Seager\inst{\ref{mitearth}, \ref{kavlimit}, \ref{mitaero} }
          \and
          Avi Shporer\inst{\ref{kavlimit}}
          \and
          Eric B. Ting\inst{\ref{nasaames}}
          \and
          R. Vanderspek\inst{\ref{kavlimit}}
          \and
          Joshua N. Winn\inst{\ref{princeton}}
}

\institute{
Instituto de Astrof\'isica de Canarias (IAC), E-38205 La Laguna, Tenerife, Spain\label{iac} \\
              \email{fmurgas@iac.es} 
\and
Departamento de Astrof\'isica, Universidad de La Laguna (ULL), E-38206 La Laguna, Tenerife, Spain\label{ull}
\and
Instituto de Astrof\'isica de Andaluc\'ia (IAA-CSIC), Glorieta de la Astronom\'ia s/n, 18008 Granada, Spain \label{iaa}
\and
NASA Exoplanet Science Institute-Caltech/IPAC, 1200 E. California Blvd. Pasadena, CA 91125, USA\label{nasaipac}
\and
Department of Physics, Engineering and Astronomy, Stephen F. Austin State University, 1936 North St, Nacogdoches, TX 75962, USA \label{austin}
\and
Cerro Tololo Inter-American Observatory, Casilla 603, La Serena, Chile \label{tololo}
\and
Department of Physics and Astronomy, The University of North Carolina at Chapel Hill, Chapel Hill, NC 27599-3255, USA\label{unc}
\and
Komaba Institute for Science, The University of Tokyo, 3-8-1 Komaba, Meguro, Tokyo 153-8902, Japan\label{utokyo1}
\and
Department of Multi-Disciplinary Sciences, Graduate School of Arts and Sciences, The University of Tokyo, 3-8-1 Komaba, Meguro, Tokyo 153-8902, Japan\label{utokyo2}
\and
Southwest Research Institute, 6220 Culebra Rd, San Antonio, TX 78238, USA\label{southwestRI}
\and
Division of Geological and Planetary Sciences, 1200 E California Blvd, Pasadena, CA, 91125, USA\label{pasadena}
\and
Department of Space, Earth and Environment, Astronomy and Plasma Physics, Chalmers University of Technology, 412 96 Gothenburg, Sweden\label{chalmers1}
\and
Department of Space, Earth and Environment, Chalmers University of Technology, Onsala Space Observatory, SE-439 92 Onsala, Sweden.\label{chalmers2}
\and
Department of Physics and Kavli Institute for Astrophysics and Space Research, Massachusetts Institute of Technology, Cambridge, MA 02139, USA\label{kavlimit}
\and
McDonald Observatory and Center for Planetary Systems Habitability\label{mcdonobs}
\and
The University of Texas, Austin Texas USA\label{utexas}
\and
Mullard Space Science Laboratory, University College London, Holmbury St Mary, Dorking, Surrey RH5 6NT, UK\label{ucl}
\and
Department of Astrophysical Sciences, Princeton University, Princeton, NJ 08544, USA\label{princeton}
\and
Department of Earth, Atmospheric, and Planetary Sciences, Massachusetts Institute of Technology, Cambridge, MA 02139, USA\label{mitearth}
\and
Department of Aeronautics and Astronautics, Massachusetts Institute of Technology, Cambridge, MA 02139, USA\label{mitaero}
\and
NASA Ames Research Center, Moffett Field, CA, 94035, USA\label{nasaames}
\and
Center for Astrophysics \textbar \ Harvard \& Smithsonian, 60 Garden Street, Cambridge, MA 02138, USA\label{harvardsmith}
\and
Dipartimento di Fisica, Universit\`a di Torino, via P. Giuria 1, I-10125 Torino, Italy\label{utorino}
\and
Th\"uringer Landessternwarte Tautenburg, Sternwarte 5, D-07778 Tautenburg, Germany\label{tautenburg}
\and
Astrobiology Center, 2-21-1 Osawa, Mitaka, Tokyo 181-8588, Japan\label{astrobiosawa}
\and
National Astronomical Observatory of Japan, 2-21-1 Osawa, Mitaka, Tokyo 181-8588, Japan\label{noao}
\and
Department of Astronomy, The Graduate University for Advanced Studies (SOKENDAI), 2-21-1 Osawa, Mitaka, Tokyo, Japan\label{sokendai}
\and
 Department of Astronomy \& Astrophysics, University of Chicago, Chicago, IL 60637, USA\label{uchicago}
\and
Institute of Planetary Research, German Aerospace Center (DLR), Rutherfordstrasse 2, D-12489 Berlin, Germany\label{dlr}
\and
Astronomical Institute of the Czech Academy of Sciences, Fri\v{c}ova 298,
25165, Ond\v{r}ejov, Czech Republic\label{czechacad}
\and
INAF – Osservatorio Astronomico di Padova, Vicolo dell’Osservatorio 5, 35122, Padova, Italy\label{inaf}
}

   \date{Received 11 July, 2022 / Accepted 2 November, 2022}

 
  \abstract
  {Ultra-short-period (USP) planets are defined as planets with orbital periods shorter than one day. This type of planets is rare, highly irradiated, and interesting because their formation history is unknown.}
   {We aim to obtain precise mass and radius measurements to confirm the planetary nature of a USP candidate found by the Transiting Exoplanet Survey Satellite (\textit{TESS}). These parameters can provide insights into the bulk composition of the planet candidate and help to place constraints on its formation history.}
   {We used \textit{TESS} light curves and HARPS-N spectrograph radial velocity measurements to establish the physical properties of the transiting exoplanet candidate found around the star HD 20329 (TOI-4524). We performed a joint fit of the light curves and radial velocity time series to measure the mass, radius, and orbital parameters of the candidate.}
   {We confirm and characterize HD 20329b, a USP planet transiting a solar-type star. The host star (HD 20329, $V = 8.74$ mag, $J = 7.5$ mag) is characterized by its G5 spectral type with $\mathrm{M}_\star= 0.90 \pm 0.05$ M$_\odot$, $\mathrm{R}_\star = 1.13 \pm 0.02$ R$_\odot$, and $\mathrm{T}_{\mathrm{eff}} = 5596 \pm 50$ K; it is located at a distance $d= 63.68 \pm 0.29$ pc. By jointly fitting the available \textit{TESS} transit light curves and follow-up radial velocity measurements, we find an orbital period of $0.9261 \pm (0.5\times 10^{-4})$ days, a planetary radius of $1.72 \pm 0.07$ $\mathrm{R}_\oplus$, and a mass of $7.42 \pm 1.09$ $\mathrm{M}_\oplus$, implying a mean density of $\rho_\mathrm{p} = 8.06 \pm 1.53$ g cm$^{-3}$. HD 20329b joins the $\sim$30 currently known USP planets with radius and Doppler mass measurements.}
   {}

   \keywords{stars: individual: HD 20329 -- planetary systems -- techniques: photometric -- techniques: radial velocities}

   \maketitle
%
\section{Introduction}

Ultra-short-period (USP) planets are defined by the community as planetary-mass objects with orbital periods shorter than one day (\citealp{Sahu2006}, \citealp{SanchisOjeda2013}). Of the $\sim$5000 exoplanets discovered to date\footnote{\url{https://exoplanetarchive.ipac.caltech.edu/}}, close to 120 are USPs, and only $\sim$30 of them have both mass and radius measurements. Some of the first discovered USP planets are Corot-7b (\citealp{Leger2009}, $P=0.85$ days), 55 Cnc e (\citealp{Dawson2010}; \citealp{Winn2011}, $P=0.74$ days), Kepler-10b (\citealp{Batalha2011}, $P=0.84$ days), and Kepler-78b (\citealp{SanchisOjeda2013}, $P=0.36$ days). The USP planets with the shortest orbital period discovered so far are the planet candidate KOI-1843.03 (\citealp{Ofir2013}) and K2-137b (\citealp{Smith2018}); these two planets revolve around M dwarfs in just 0.18 days ($\sim$4.3 hours). 

The first systematic search for transiting USPs was conducted by \citet{SanchisOjeda2014}. Using \textit{Kepler} data, they found 106 USP candidates, 6 of which have $P<6$ hours. The authors noted that the planetary radii of these objects were rarely larger than $2.0$ $R_\oplus$. \cite{Winn2018} also found that the radius distribution of USPs declines sharply around 2 $\mathrm{R}_\oplus$, and they proposed that this sharp decline is attributed to photoevaporation of the planet atmosphere that is caused by radiation from the host star.

Although the period cutoff ($P < 1$ day) is arbitrary, this type of objects appears to have some distinct characteristics that may indicate that they are a distinct population. For example, the occurrence rates of USP planets seem to depend on spectral type. \citet{SanchisOjeda2014} found that the USP occurrence rate falls from $1.1 \pm 0.4$ \% for M dwarfs to $0.15 \pm 0.05$ \% for K dwarfs. In the same work, they establish that the occurrence rate for G dwarfs is $0.51 \pm 0.07$ \%, meaning that about 1 in 200 solar-type stars hosts a USP planet (for planets larger than 0.84 $\mathrm{R}_\oplus$).

There are other characteristics that separate USP planets from other types of short-period exoplanets such as hot Jupiters. \cite{Winn2017} found that the metallicity distribution for stars that host hot Jupiters is different from that of stars that host USP planets: Hot-Jupiter host stars tend to be more metal rich than the stars that have USP planets. \cite{Winn2018} also pointed out that, unlike hot Jupiters, USP planets can be found in multiple planet systems. When USP planets are found in multiple planet systems, the ratio of the orbital periods between the inner planet and its nearest neighbor is typically higher than 4 ($P_2/P_1 \gtrsim 4$) (\citealp{Steffen2013}, \citealp{Winn2018}, \citealp{Pu2019}). This period ratio is higher than what is seen in \textit{Kepler} multiplanet systems (\citealp{Fabrycky2014}). Additionally, for systems of multiple transiting planets, the dispersion of the orbital inclination of the different transiting planets has been found to be higher when a USP planet is part of the system (\citealp{Dai2018}).

The origin of USP planets is still an open issue. Their close-in orbit usually places them inside the dust-sublimation region around their host star, meaning that they are unlikely to have formed in their present-day orbits. Hence, in all the proposed scenarios, these objects formed in a wider orbit around their host star and migrated to their current position. \cite{Petrovich2019} proposed that secular interactions in multiplanet systems (e.g., $N_{\mathrm{planets}} > 3$) can affect the inner planet (with $P$ in the 5-10 day range) and push it into a highly eccentric orbit. It is eventually tidally captured by the star in a short-period orbit that is circularized over time as a result of planetary tides. \cite{Pu2019} explored another formation mechanism in which a small rocky planet is born with a period of a few days and moderate eccentricity ($e \gtrsim 0.1$) in a multiplanetary system; the outer planets tidally interact with each other and with the innermost planet, damping the eccentricity to a value close to zero and shrinking the semi-major axis in a quasi-equilibrium state. At the end of this process, the innermost planet becomes a USP planet, while the second planet, an Earth- or super-Earth-sized planet, stabilizes itself on a 10 day orbit. TOI-500 is the first four-planet system for which this mechanism has been proven to work \citep{Serrano2022}. \cite{Millholland2020} proposed an obliquity-driven tidal migration mechanism. There, in a system of planets with strong mutual inclinations, planetary obliquities and tides become excited in a positive-feedback loop that forces inward migration, until a condition is reached in which the high obliquities are tidally destabilized and migration stalls.

Because USP planets are rare, those with well-determined parameters are especially useful for comparison with the predictions made by theoretical models. The NASA-sponsored Transiting Exoplanet Survey Satellite (\textit{TESS}; \citealp{Ricker2014}) is a space telescope equipped with four cameras observing an area of sky of $24^\circ \times 96^\circ$ degrees. Its main goal is to search for transiting exoplanets around bright stars ($5 < I_C < 13$ mag) by observing the same part of the sky almost uninterrupted during $\sim$27 days. Launched in April 2018, \textit{TESS} is past its original two-year mission and is currently in its extended mission, which was approved to continue until the end of September 2022. Although USPs are rare, the \textit{TESS} $\sim$27-day observation cycles offer a unique opportunity to find this type of objects, and because the mission focused on bright stars, it allows Doppler mass measurements through radial velocity follow-up observations.

Here we report the discovery of HD 20329b, a USP planet around a bright ($V = 8.74$ mag, $J = 7.5$ mag) G-type star, which was discovered using \textit{TESS} data. This paper is organized as follows: in Section \ref{sec:observations} we describe the \textit{TESS} data, spectroscopic follow-up, and high resolution imaging of HD 20329. In Section \ref{sec:analysis} we describe the methods we used to determine the stellar parameters, the light curve, and the radial velocity fitting procedure. In Section \ref{sec:results} we present the parameters of HD 20329b and place this planet in the context of known USP planets. Section \ref{sec:conclusions} presents the conclusions of this work.

\section{Observations}
\label{sec:observations}
\subsection{\textit{TESS} photometry}

HD 20329 (TIC 333657795; \citealp{Stassun2018}) was observed by \textit{TESS} from 20 August 2021 until 16 September 2021 (sector 42) and from 16 September 2021 until 12 October 2021 (sector 43). For the stellar coordinates and magnitudes, see Table \ref{Table:StarInfo}. For sector 42, the target was observed on camera 4 CCD 3, while in sector 43, the star was placed on camera 2 CCD 1. For each \textit{TESS} sector, the star was observed for $\sim$25 days, and the images were stacked using a 2-minute cadence mode.

\textit{TESS} observes a field continuously for $\sim$27 days. At every orbit perigee ($\sim$13 days), science operations are interrupted, and the data are sent to Earth for processing. The raw images were processed by the Science Processing Operations Center (SPOC) at NASA Ames Research Center. The SPOC pipeline (\citealp{Jenkins2016}) performs image calibration and data-quality control, extracts photometry for all the \textit{TESS} target stars in the field of view, and searches the extracted light curves for transit signatures.

The data reduction process of \textit{TESS} time series starts by using simple aperture photometry (SAP: \citealp{Morris2020}) to generate an initial light curve. Then the Presearch Data Conditioning (PDC) pipeline module (\citealp{Smith2012}, \citealp{Stumpe2014}) removes some instrumental systematic effects from the time series. Transit events are searched with the wavelet-based matched filter described in \cite{Jenkins2002} and \cite{JenkinsJM2020}, and are then fit to transit models, including the contribution made by stellar limb-darkening effects (\citealp{Li2019}). Finally, a set of diagnostic tests are applied to the light curves to establish whether the detected transit events have a planetary origin (\citealp{Twicken2018}). After this process, the \textit{TESS} science office reviews the transit signature and promotes the candidate as a \textit{TESS} object of interest (TOI) if it has a likely planetary origin. In the case of HD 20329, the planetary candidate was assigned the TOI identification TOI-4524.01, and the community was alerted to it in October 2021. The SPOC transit depth and orbital period of TOI-4524.01 were $210 \pm 0.60$ ppm and $P=0.926014 \pm 0.00005$ days, making this TOI a USP planet candidate.

The \textit{TESS} light curves were analyzed independently using the D\'etection Sp\'ecialis\'ee de Transits (DST; \citealp{Cabrera2012}) pipeline. Variability in the PDC-SAP light curve was first removed using a Savitzky-Golay filter \citep{1964AnaCh..36.1627S,1992nrca.book.....P}, and transit searches were performed. A transit signal with an orbital period of $0.92653 \pm 0.00011$ day and a transit depth of $204 \pm 17$ ppm was detected, consistent with the signal detected by SPOC.

 \begin{figure*}
   \centering
   \includegraphics[width=\hsize]{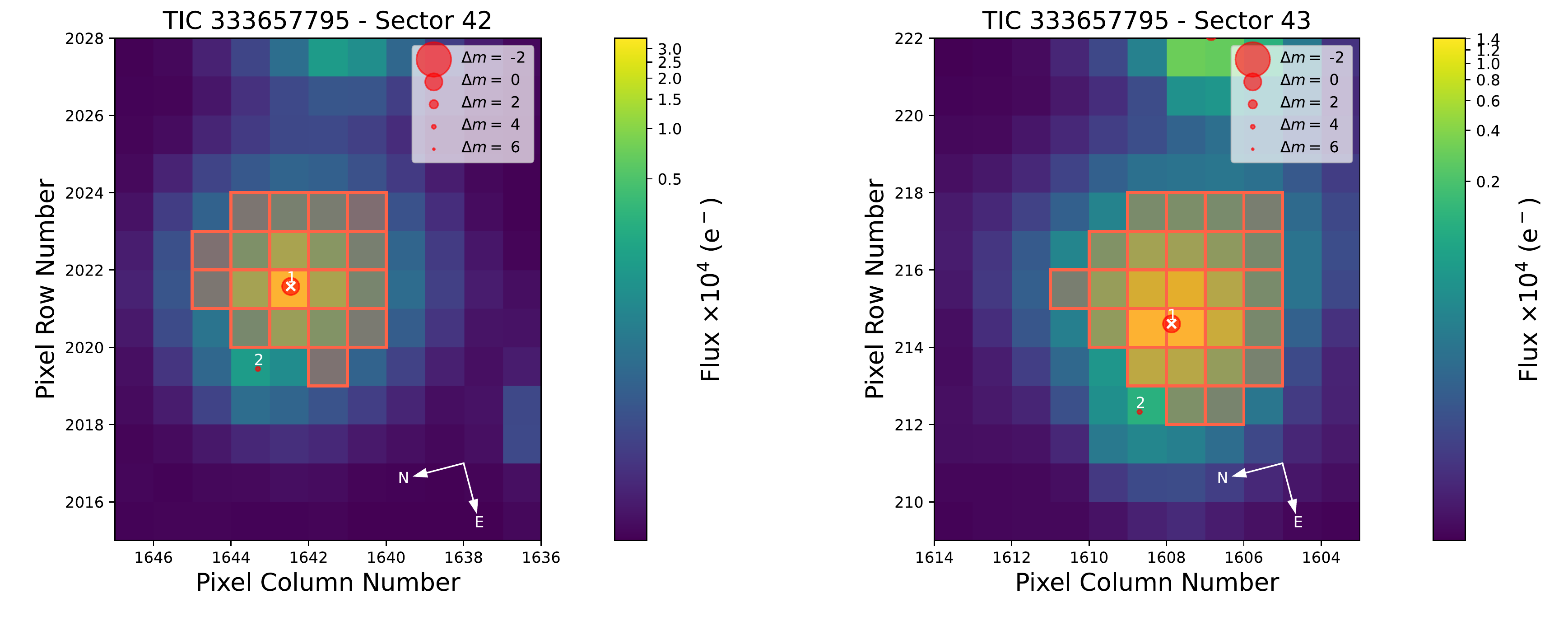}
   \caption{\textit{TESS} target pixel file image of HD 20329 observed in sector 42 (left) and sector 43 (right), made with \texttt{tpfplotter} (\citealp{Aller2020}). The pixels highlighted in red show the aperture used by \textit{TESS} to create the light curves. The position and sizes of the red circles represent the position and \textit{TESS} magnitudes of nearby stars, respectively.}
   \label{Fig:TESS_TPF_S42andS43}
\end{figure*}

We used the \textit{TESS} PDC-SAP light curves for the transit analysis presented in this work. The PDC-SAP curves were corrected for instrumental systematic effects and include some correction for flux contamination from nearby stars. The \textit{TESS} sector 42 and 43 observations are publicly available at the Barbara A. Mikulski Archive for Space Telescopes (MAST\footnote{\url{https://mast.stsci.edu/portal/Mashup/Clients/Mast/Portal.html}}). Figure \ref{Fig:TESS_TPF_S42andS43} shows the \textit{TESS} target pixel file (TPF) images around HD 20329 for sectors 42 and 43, and the photometric aperture used by \textit{TESS} is highlighted with red squares.

\begin{table*}
  \caption[]{HD 20329 identifiers, coordinates, stellar parameters, and magnitudes.}
  \label{Table:StarInfo}
  \centering
  \begin{tabular}{llcc}
    \hline \hline
    Identifiers & & & Ref. \\
    \hline
    HD & & 20329 & \\
    HIP & &  15249 & \\
    TIC & & 333657795 & \\
    TOI & & 4524 & \\
    2MASS & & J03164262+1539260 & \\
    \textit{Gaia} EDR3 & & 30398648945512960 & \\
    \hline
    Equatorial coordinates & & & \\
    \hline
    RA (J2000) & & $03^{\mathrm{h}}\; 16^{\mathrm{m}}\; 42^{\mathrm{s}}.63$ & 1 \\
    DEC (J2000) & & $+15^{\circ}\; 39'\; 26\farcs01 $ &  1 \\ 
    $\mu_{\mathrm{RA}}$ [mas/year] & & $111.77 \pm 0.022$ & 1 \\
    $\mu_{\mathrm{DEC}}$ [mas/year] & & $-202.41 \pm 0.018$ & 1 \\
    Parallax [mas] & & $15.66 \pm 0.02$ & 1 \\
    Distance [pc] & & $63.68^{+0.29}_{-0.28}$ & 2 \\
    Star systemic radial velocity [km/s] & & $-71.539 \pm 0.005$ & 6 \\
    \hline
    Stellar Parameters & & & \\
    \hline
    Effective Temperature [K] & $\mathrm{T}_{\mathrm{eff}}$ & $5596 \pm 50$ &  6 \\
    Stellar Luminosity [$\mathrm{L}_\odot$] & $\mathrm{L}$ & $1.12 \pm 0.006$ & 6 \\
    Surface gravity [cm/s$^2$] & $\log(g)$ & $ 4.40 \pm 0.07$ & 6 \\
    Metallicity [dex] & $[\mathrm{Fe}/\mathrm{H}]$ & $ -0.07 \pm 0.06$ & 6 \\
    Activity index & $\log R'_{HK}$ & $-5.03 \pm 0.03$ & 6 \\
    Stellar Age [Gyr] & & $ 11 \pm 2$ &  6 \\
    Projected stellar rotational velocity [km/s] & $v\sin(i)$ & $ 3.5 \pm 0.6$ &  6 \\
    Microturbulence velocity [km/s] & $v_t$ & $ 0.86 \pm 0.04$ &  6 \\
    Mass [M$_\odot$] & $\mathrm{M}_\star $ & $ 0.90 \pm 0.05 $ & 6 \\
    Radius [R$_\odot$] & $\mathrm{R}_\star $ & $ 1.13 \pm 0.02 $  & 6 \\
    Derived stellar density [g/cm$^3$] & $\rho_\star $ & $ 0.88 \pm 0.068 $ & 6 \\
    \hline
    Apparent magnitudes & & & \\
    \hline
    \textit{Gaia} G [mag] & & $ 8.600 \pm 0.003$ & 1 \\
    B [mag] & & $9.466 \pm 0.017$ & 3 \\
    V [mag] & & $8.738 \pm 0.026$ & 3 \\
    Sloan g [mag] & & $9.057 \pm 0.017$ & 3 \\
    Sloan r [mag] & & $8.573 \pm 0.029$ & 3 \\
    Sloan i [mag] & & $8.432 \pm 0.022$ & 3 \\
    J [mag] & & $7.492 \pm 0.021$ & 4 \\
    H [mag] & & $7.208 \pm 0.049$ & 4 \\
    K [mag] & & $ 7.116 \pm 0.024$ & 4 \\
    WISE W1 [mag] & & $7.054 \pm 0.041$ & 5 \\
    WISE W2 [mag] & & $7.105 \pm 0.020$ & 5 \\
    WISE W3 [mag] & & $7.120 \pm 0.017$ & 5 \\
    WISE W4 [mag] & & $7.111 \pm 0.108$ & 5 \\
    \hline
  \end{tabular}
  
\tablebib{ (1)~\citet{GaiaEDR32020}; (2) \citet{BailerJones2018} ; (3) \citet{Henden2015}; (4) \citet{Cutri2003}; (5) \citet{Cutri2014}; (6) This work.}
\tablefoot{The parameters for HD 20329 from \textit{Gaia} EDR3 did not change with the recent release of \textit{Gaia} DR3 (\citealp{GaiaDR32022}).}
  
\end{table*}
  
\subsection{Ground-based seeing-limited photometry with MuSCAT2}
\textit{TESS} has a relatively large pixel scale (21 $\arcsec$/pixel), hence the detected transit event might originate from another star located inside the \textit{TESS} photometric aperture. Centroid analysis results from the \textit{TESS} pipeline using images from sectors 42 and 43 combined eliminate the possibility that nearby catalog stars caused the transit signature. In particular, they exclude the 17th$^{\mathrm{}}$ mag star 37$\arcsec$ southwest of HD 20329, the 12th$^{\mathrm{}}$ mag star 49$\arcsec$ northwest of it, as well as some very dim stars slightly over 25$\arcsec$ to the south-southwest. 

When \textit{TESS} announced HD 20329 as an object of interest, the transit depth found by the pipeline was $\sim$0.230 mmag (210 ppm), meaning that the transit event probably cannot be detected on HD 20329 with the medium-sized ground-based telescopes that are typically used to rule out false positives. Nonetheless, it is possible to rule out other stars in the field as causing the transit signal based on seeing-limited photometry.

HD 20329 was observed on the nights of 4 and 5 December 2021 UT with the simultaneous multicolor imager MuSCAT2 \citep{Narita2019} mounted on the 1.5 m Telescopio Carlos S\'{a}nchez (TCS) at Teide Observatory, Spain. MuSCAT2 has four CCDs with $1024 \times 1024$ pixels, and each camera has a field of view of $7.4\arcmin \times 7.4 \arcmin$ (pixel scale of 0.44 $\arcsec$/pixel). The instrument is capable of taking images simultaneously in $g'$, $r'$, $i'$, and $z_s$ bands with short read-out times. For both nights, the raw data were reduced by the MuSCAT2 pipeline \citep{Parviainen2019}, which performs standard image calibration and aperture photometry and is capable of modeling the instrumental systematics present in the data while simultaneously fitting a transit model to the light curve.

On the night of 4 December 2021 UT, the telescope was defocused to avoid saturation of HD 20329. On the night of 5 December 2021 UT, we also defocused the telescope, but we let the target star saturate in order to detect fainter stars in the field. For both nights, the exposure times were set to 5 seconds for all the MuSCAT2 bands. For the first night, we were unable to detect the transit on target, likely due to the small transit depth of the event; on the second night HD 20329 was saturated on purpose so that no transit detection could be performed. Inside a field of view with a radius of 2.5\arcmin , four stars identified by \textit{Gaia} (DR3; \citealp{GaiaDR32022}) can produce a transit signal with the depth detected by \tess. Of these stars, we were only able to rule out the nearby star (TIC 333657797, $V = 12.5$ mag) as an eclipsing binary causing the transit events. The other three \textit{Gaia} sources were too faint and are not detected in the MuSCAT2 images.

\subsection{Ground-based high-resolution imaging}
\label{sec:HighResImg}
Part of the validation process of transiting exoplanets is the assessment of possible flux contamination by nearby companions (bound or unbound to the target star) and its effect on the derived planetary radius \citep{ciardi2015}. For this reason, we observed HD~20329 with a combination of high-resolution imaging resources including optical speckle and near-infrared (NIR) adaptive optics (AO). Astrometric data from \textit{Gaia} EDR3 (\citealp{GaiaEDR32020}) was also used to provide additional constraints on the presence of undetected stellar companions as well as wide companions.

\subsubsection{SOAR optical speckle imaging}
\begin{figure}
   \centering
   \includegraphics[width=\hsize]{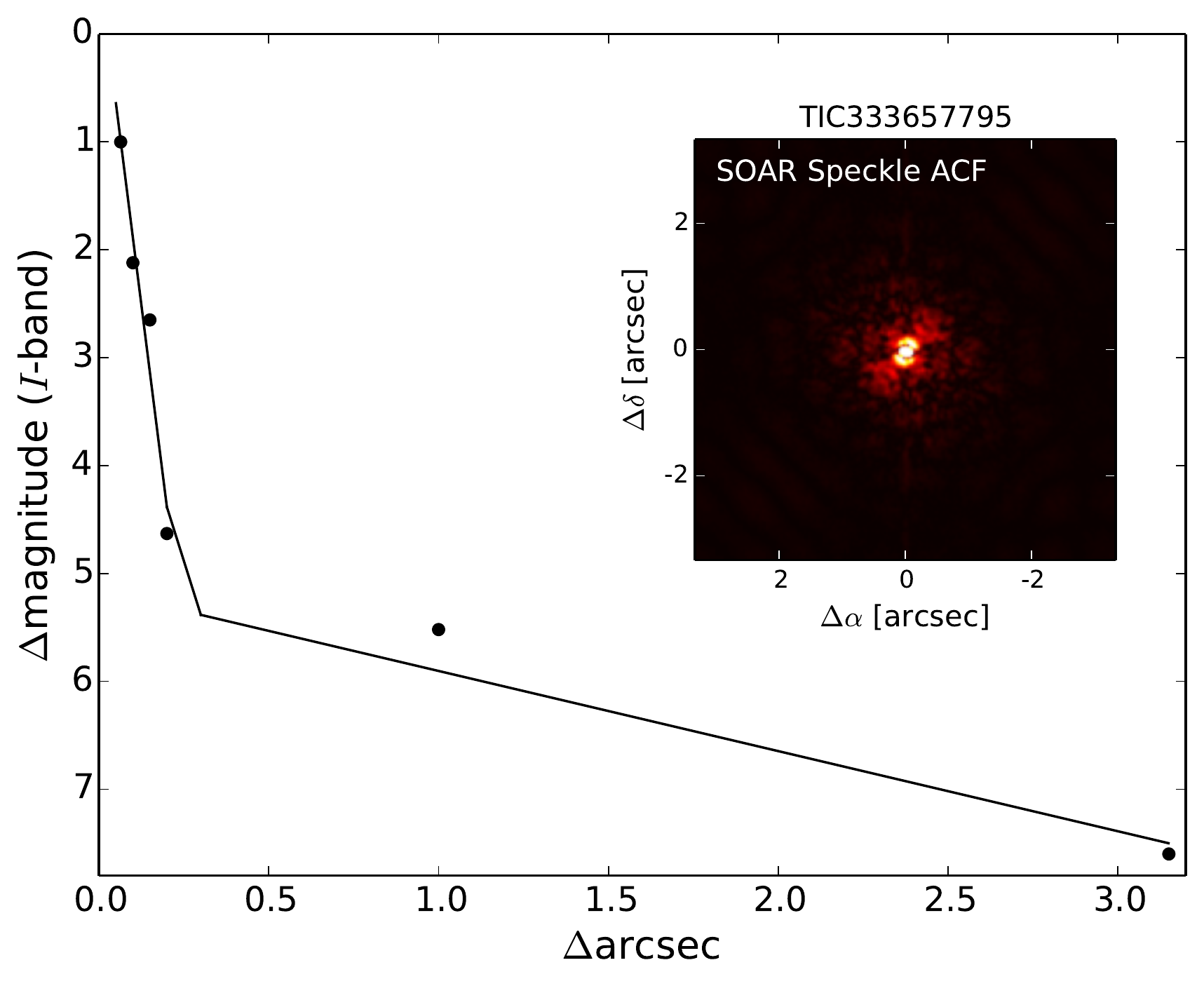}
   \caption{Speckle sensitivity curve and ACF (inset) from SOAR HRCam observations of HD 20329. No nearby stars were detected within 3\arcsec of HD 20329 with HRCam.}
   \label{Fig:AO_SOAR}
 \end{figure}

We observed HD 20329 using speckle imaging on 20 November 2021 UT with the 4.1 m Southern Astrophysical Research (SOAR) telescope (\citealp{Tokovinin2018}). The observations were made as part of the SOAR \textit{TESS} survey (see \citealp{Ziegler2020} for details). The data were acquired with the HRCam instrument (Cousins I band). The observation reached a sensitive threshold to ensure the detection of stars 5.8 magnitudes fainter than the target at an angular distance of 1$\arcsec$. No nearby stars around HD 20329 were detected using a search radius of 3$\arcsec$. Figure \ref{Fig:AO_SOAR} shows the 5$\sigma$ detection sensitivity and speckle autocorrelation functions (ACF) from the SOAR observations.

\subsubsection{Palomar NIR AO imaging}
\begin{figure}
   \centering
   \includegraphics[width=\hsize]{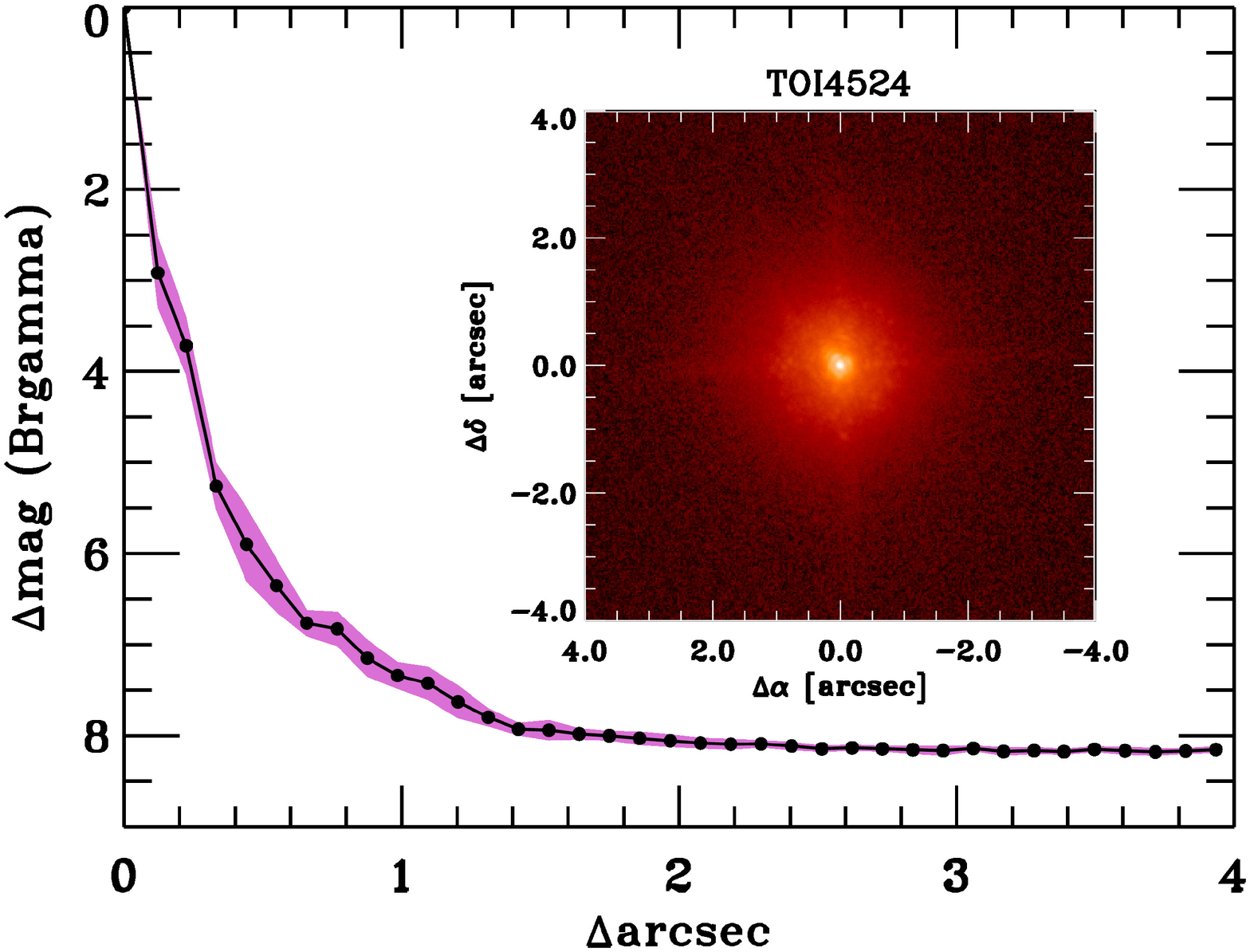}
   \caption{Palomar NIR AO imaging and sensitivity curves for HD~20329 taken in the Br$\gamma$ filter. The images were taken in good seeing conditions, and we reach a contrast of 7 magnitudes fainter than the host star within 0.5\arcsec. {\it Inset:} Image of the central portion of the data, centered on the star.
    }
   \label{Fig:AO_Palomar}
 \end{figure}

HD 20329 was observed by the adaptive-optics instrument PHARO (\citealp{hayward2001}; field of view of $\sim$ $25\arcsec$, pixel scale of $0.025\arcsec$ per pixel) mounted on the 5 m Hale telescope on 11 November 2021. The observations were made using the narrow-band NIR $Br-\gamma$ filter and applying a five-point dither pattern. The raw data were reduced using standard procedures (flat, dark, and sky calibration; calibrated science images were coadded). The 5$\sigma$ sensitive curve presented in Figure~\ref{Fig:AO_Palomar} was determined using injection of artificial sources following the procedure outlined in \citep{furlan2017}. No stellar companions were detected in PHARO observations.

\subsubsection{\textit{Gaia} assessment}
We used \textit{Gaia} EDR3 measurements to search for stellar companions around HD 20329 following \citet{Luque2022} (see section 3.2.3 of \citealp{Luque2022} and references therein). There are no nearby stars around the position HD 20329 with similar astrometric properties (i.e., parallaxes and proper motions; \citealp{mugrauer2020,mugrauer2021}) that could indicate that they are bound to the host star. Additionally, the \textit{Gaia} renormalized unit weight error (RUWE), a metric used to measure the level of astrometric noise caused by a gravitationally bound unseen companion \citep[e.g., ][]{Ziegler2020}, is consistent with a single-star model for the case of HD 20329 (EDR3 RUWE = 0.92, below the $1.4$ RUWE threshold for a multiple star system). 

\subsection{Spectroscopic observations}
Between 29 November 2021 (UT) and 30 January 2022 (UT), we collected 120 spectra with the High Accuracy Radial velocity Planet Searcher for the Northern hemisphere \citep[HARPS-N: $\lambda$\,$\in$\,(378--691)\,nm, R$\approx$115\,000,][]{2012SPIE.8446E..1VC} mounted at the 3.58 m Telescopio Nazionale Galileo (TNG) of Roque de los Muchachos Observatory in La Palma, Spain, under the observing program CAT21A\_119. The exposure time was set to 259--1800 seconds, based on weather conditions and scheduling constraints, leading to a S/N per pixel of 20--136 at 5500\,\AA. The spectra were extracted using the off-line version of the HARPS-N data reduction software (DRS) pipeline \citep{2014SPIE.9147E..8CC}, version 3.7. Doppler measurements and spectral activity indicators (CCF\_BIS, CCF\_FWHM, CCF\_CTR and Mont-Wilson S-index, S\_{MW}) were measured using an online version of the DRS, the YABI tool\footnote{Available at \url{http://ia2-harps.oats.inaf.it:8000}.}, by cross-correlating the extracted spectra with a G2 mask \citep{1996A&AS..119..373B}. We also used {\tt serval} code (\citealp{2018A&A...609A..12Z}) to measure relative RVs by the template-matching, chromatic index (CRX), differential line width (dLW), and H$\alpha$, sodium Na~D1 \& Na~D2 indexes. The uncertainties in the relative RVs measured with {\tt serval} are in the range 0.5--4.4\,\mps, with a mean value of 1.2\,\mps. The uncertainties in the absolute RVs measured with the online version of DRS (YABI) are in a the range 0.6--6.1\,\mps, with a mean value of 1.3\,\mps. Table~\ref{table-TOI-4524-tng_harpn-0120-drs-complete_output} gives the time stamps of the spectra in BJD$_{\mathrm{TDB}}$, absolute RVs, and spectral activity indicators (CCF\_BIS, CCF\_FWHM, CCF\_CTR, and Mont-Wilson S-index, S\_{MW}) measured with YABI. Table~\ref{table-TOI-4524-tng_harpn-0120-srv-complete_output} gives the relative RVs and spectral activity measurements (CRX, dLW, and H$\alpha$, and sodium Na~D1 \& Na~D2 indexes) measured with {\tt serval}. In the joint RV and transit analysis presented in Section~\ref{sec:transit_and_rv_fit}, we used relative RVs measured from HARPS-N spectra with {\tt serval} by the template-matching technique.

\begin{figure} 
\begin{center}
 \includegraphics[width=0.5\textwidth]{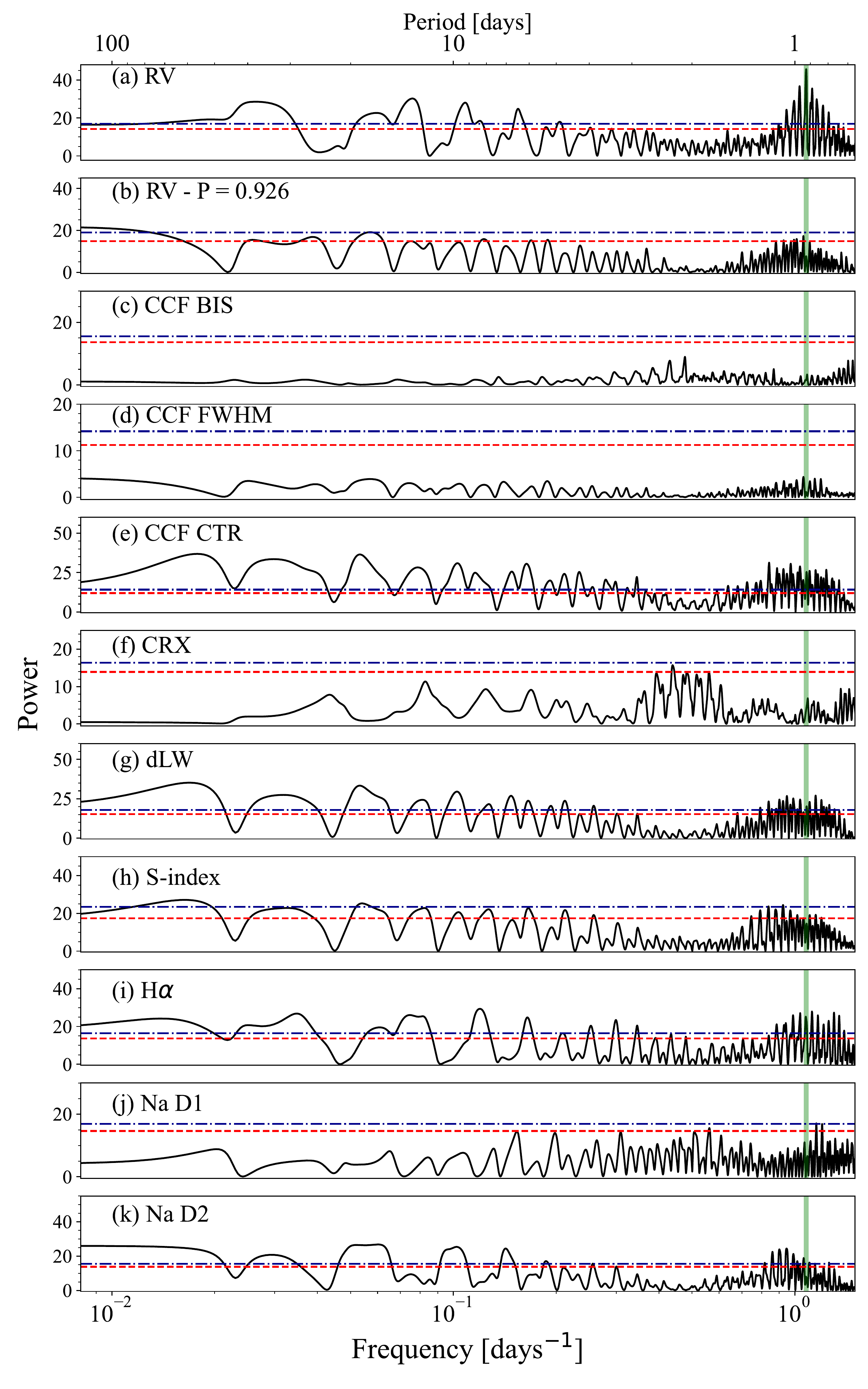}
\end{center}
\caption{GLS periodogram (\citealp{Zechmeister2009}) for HARPS-N radial velocity measurements. (a) Radial velocity measurement residuals after fitting the transiting-planet signal. (b) Stellar activity indices of HD 20329 ((c) to (k)). The peak with the maximum power in the periodogram for the RV measurements corresponds to a period of $P=0.9263$ days (vertical green line), in agreement with the period of the transiting-planet candidate reported by \textit{TESS}. The dash-dotted blue and segmented red line represent the 0.01\% and 0.1\% FAP levels, respectively.}
\label{Fig:HARPS_Periodogram}
\end{figure}

Figure \ref{Fig:HARPS_Periodogram} shows the generalized Lomb-Scargle (GLS; \citealp{Zechmeister2009}) periodogram for the radial velocities and activity indices. The false-alarm probability (FAP) levels were computed using the bootstrap method (\citealp{Murdoch1993}, \citealp{Hatzes2016}) with 10000 iterations. The peak of the RVs GLS periodogram is located at $P=0.9263$ days, in agreement with the orbital period of the transiting candidate announced by \textit{TESS}.

\section{Analysis}
\label{sec:analysis}
\subsection{Stellar parameters}
\label{sec:stellar_param}
The stellar parameters for HD 20329 are presented in Table \ref{Table:StarInfo}. The analysis of the coadded HARPS-N stellar spectrum was carried out by using the \texttt{BACCHUS} code \citep{2016ascl.soft05004M}, which relies on the MARCS model atmospheres \citep{2008A&A...486..951G} and atomic and molecular line lists from \citet{2021A&A...645A.106H}. In brief, the surface gravity was determined by requiring ionization balance of the \ion{Fe}{I} lines and \ion{the Fe}{II} line. A microturbulence velocity  was also derived by requiring no trend of Fe line abundances against their equivalent widths. The output metallicity  is represented by the average abundance of the \ion{Fe}{I} lines. An effective temperature of 5596\,$\pm$\,50\,K was derived  by requiring no trend of the \ion{Fe}{I} lines abundances against their respective excitation potential.

We compared our stellar parameters to those found in the literature in the Hypatia Catalog \citep{Hinkel14}. HD 20329 was spectroscopically observed by \citet{Ramirez13} and \citet{2017ApJ...839...94B}. The effective temperatures and iron-content determined here are in excellent agreement with both \citet{Ramirez13} and \citet{2017ApJ...839...94B}, especially when [Fe/H] is normalized to the same solar scale. While the surface gravities of \citet{Ramirez13} and \citet{2017ApJ...839...94B} disagree within their own errors, the value that we determined (4.40 $\pm$ 0.07) overlaps with the two literature determinations within errors. The [Fe/H] of HD 20329, in addition to other iron-peak elements (Cr, Mn, and Ni), is somewhat subsolar. However, many elements, especially the $\alpha$-elements (C, O, Si, Ca, Ti), within HD 20329 are supersolar \citep{Ramirez13, 2017ApJ...839...94B}. In combination with these abundance trends, the stellar kinematics indicate that HD 20329 likely originated from the thick disk according to a conservative kinematic prescription by \citet{Bensby03} and as noted within \citet{2017ApJ...839...94B}.

The stellar rotation ($v \sin i$) was estimated by measuring the average of the Fe line broadening after subtracting the instrument and the thermal, collisional, and microturbulence broadening. This velocity leads to an estimate of the rotational period of 15 $\pm$ 3 days (assuming $\sin i = 1$). Based on the \citet{1984ApJ...279..763N} and \citet{2008ApJ...687.1264M} activity-rotation relations and using (B-V) of 0.670 and the \logrhk{} measured with YABI (-5.03\,$\pm$\,0.03), we estimated a rotation period of \host{} for 28.6\,$\pm$\,5.8\,days and 30.7\,$\pm$\,3.3\,days, respectively. The discrepancy between the two indicators may be explained by the fact that for relatively long rotation periods, its impact on the line broadening becomes very weak and its disentanglement from other sources of broadening becomes more uncertain. Although the empirical calibrations of chromospheric activity and rotational periods have their own issues, we nevertheless favor a period of $\sim$30 days. We tried to establish the rotation period of HD 20329b using long-term photometry data from several surveys. We were only able to access the photometry from the ASAS-SN public light-curve archive\footnote{\url{https://asas-sn.osu.edu/}} (\citealp{Shappee2014}, \citealp{Kochanek2017}), but found no conclusive signs of photometric modulation attributable to the rotation period of the star (see Appendix \ref{Appendix:StellarRot}).   

In a second step, we used the Bayesian tool PARAM \citep{2014MNRAS.445.2758R,2017MNRAS.467.1433R} to derive the stellar mass, radius, and age using the spectroscopic parameters and the updated \textit{Gaia} EDR3 luminosity along with our spectroscopic temperature. The resulting error radius is shown in Table~\ref{Table:StarInfo} and appears to be particularly small (1.8\%). The very precise luminosity provided by Gaia allows us to constrain stellar parameters such as the radius very well. However, these Bayesian tools underestimate the error budget because they do not take the systematic errors between one set of isochrones to the next into account because of the various underlying assumptions in the respective stellar evolutionary codes. In order to attempt to take these systematic errors into account, we combined the results of the two sets of isochrones provided by PARAM (i.e., MESA and Parsec) and added the difference between the two sets of results to the error budget provided by PARAM. However, although using two set of isochrones may mitigate underlying systematic errors, our formal error budget for radius and luminosity maybe still be underestimated, as demonstrated by \citet{Tayar2022}. For solar-type stars such as HD 20329, absolute errors may rather be up to 4\%, 2\%, 5\%, and 20\% for radius, luminosity, mass, and age, respectively.

Despite its nearly solar metallicity, the derived age from the isochrones indicates that the star is old (11 Gyr). In parallel, the HARPS-N spectrum allows us to check other indices of age, notably the chromospheric activity of the Ca H\&K lines and the Li abundance. We did not observe any sign of chromospheric activity in the cores of the Ca H and K lines (hence leading to a relatively high \logrhk{}). Using the activity-age relation of \citet{2008ApJ...687.1264M}, we also found the age of \host{} to be in a range of 4--8\,Gyr, which is quite consistent with the age of 9--13\,Gyr as determined by isochrone fitting. In addition, we did not observe any lithium line, which supports the idea that the old age of the stars permits a slow depletion of all the lithium in its shallow convective external envelope.

\subsection{Transit light curve and radial velocity model fit}
\label{sec:transit_and_rv_fit}
To obtain the planetary mass and radius, we fit the light curves and radial velocity measurements simultaneously following the procedure described in \citet{Murgas2021}. In summary, the transits were modeled with \texttt{PyTransit}\footnote{\url{https://github.com/hpparvi/PyTransit}} (\citealp{Parviainen2015}) adopting a quadratic limb-darkening (LD) law. The fitted LD coefficients ($\mathrm{u}_1$ and $\mathrm{u}_2$) were weighted against the predicted values computed by \texttt{LDTK}\footnote{\url{https://github.com/hpparvi/ldtk}} (\citealp{Parviainen2015b}) while using the \cite{Kipping2013} coefficient parameterization. The HARPS-N radial velocity measurements were fit using \texttt{RadVel}\footnote{\url{https://github.com/California-Planet-Search/radvel}} (\citealp{Fulton2018}). For the transits, we set as free parameters the planet-to-star radius ratio $\mathrm{R}_\mathrm{p}/\mathrm{R}_\star$, the quadratic LD coefficients, the central time of the transit $\mathrm{T}_{\mathrm{c}}$, the planetary orbital period $P$, the stellar density $\rho_\star$, the transit impact parameter $\mathrm{b}$, and in the case of noncircular-orbit models, the eccentricity $e$ and argument of the periastron $\omega$. The free parameters for the radial velocity fit were the radial velocity semi-amplitude ($K_\mathrm{RV}$), the host star systemic velocity ($\gamma_0$), and the instrumental jitter ($\sigma_{\mathrm{RV\; jitter}}$). For the RV modeling, the orbital period, central transit time, eccentricity, and argument of the periastron were also set free, but were taken to be global parameters in common with the transit model. We also introduced a term ($\dot{\gamma}$) to model the slope seen in the RV measurements (see Fig. \ref{Fig:HARPSN_RVs} top and middle panel).

As a result of their short orbital periods, it is expected that the orbits of USPs become circularized over time. Nevertheless, we decided to fit a noncircular orbit to the RVs, setting as global free parameters the eccentricity and the argument of the periastron using the parameterization $\sqrt{e}\sin(\omega)$ and $\sqrt{e}\cos(\omega)$ (parameter limits $[-1,1]$). With this parameterization, we sample values of $e \in [0,1]$ and $\omega \in [0, 2\pi]$.

We modeled the residual correlated noise in \textit{TESS} and HARPS-N data using Gaussian processes (GPs; e.g., \citealp{Rasmussen2006}, \citealp{Gibson2012}, \citealp{Ambikasaran2015}). For \textit{TESS} light curves, we chose a Matern $3/2$ kernel,

\begin{equation}
    k_{ij\; \mathrm{TESS}} = c^2_1 \left( 1 + \frac{\sqrt{3} |t_i-t_j|}{\tau_1}\right) \exp\left(-\frac{\sqrt{3} |t_i-t_j|}{\tau_1}\right)
\label{Eq:TESS_GPKernel}
,\end{equation}
where $|t_i-t_j|$ is the time between epochs in the series, and the hyperparameters $c_1$ and $\tau_1$ were set free. Because of its flexibility, the Matern $3/2$ kernel is a commonly used kernel for modeling \textit{TESS} systematics.

The radial velocity package \texttt{RadVel} allows modeling correlated noise using GPs. For the radial velocity time series, we chose an exponential squared kernel (i.e., a Gaussian kernel),

\begin{equation}
    k_{ij\; \mathrm{RV}} = c^2_2 \exp \left(  - \frac{ (t_i-t_j)^2}{\tau^2_2} \right)
\label{Eq:RV_GPKernel}
,\end{equation}
where $t_i-t_j$ is the time between epochs in the series, and the hyperparameters $c_2$ and $\tau_2$ were set free. We chose a Gaussian kernel to model the red noise in the RVs since it is a simple model and we saw no signs of RV signals induced by the rotation period of the star in our RV measurements periodogram, nor evident correlations between the RVs and the measured activity indices. This is probably caused by the fact that we likely covered only two rotation periods of the star (RV baseline of 62 days) if the stellar rotation is around 30 days based on activity relations.

For the fitting procedure, we constrained the prior values for the orbital period and central time of the transit based on the results of applying \texttt{Transit Least Squares} (\texttt{TLS}, \citealp{Hippke2019}) to the \textit{TESS} time series. Then we maximized a posterior function for the joint data set (\textit{TESS} plus HARPS-N) using \texttt{PyDE}\footnote{\url{https://github.com/hpparvi/PyDE}}, and started a Markov chain Monte Carlo (MCMC) using \texttt{Emcee} (\citealp{ForemanMackey2013}) to explore the parameter space. We used 160 chains to fit 19 free parameters, 2000 iterations as a burn-in phase, and 8000 iterations for the main MCMC phase. After the MCMC was complete, we computed the percentiles corresponding to the median and 1$\sigma$ limits (from the median) of the distribution for each variable. We adopted these values as the final parameter estimates and their corresponding uncertainties. 

We repeated this procedure to test whether the data supported a circular or an eccentric model and the use of GPs for modeling the RVs. We compared four models: 1) a circular-orbit model without GPs for the RV data, 2) a circular-orbit model including GPs (for \textit{TESS} and the RVs), 3) a noncircular-orbit model using GPs only for \textit{TESS} photometry, and 4) a noncircular-orbit fit including GPs (for \textit{TESS} and HARPS-N RVs). Then we computed the model comparison metric Bayesian information criterion (BIC; \citealp{Schwarz1978}) to determine the best approach to fitting our data. We found that the preferred model based on these criteria was the circular-orbit model including GPs for both \textit{TESS} and RV measurements, with a BIC difference between models of $\Delta \mathrm{BIC} = -7.6$ when compared to the second best model, that is, the circular model without GPs for the RVs (see Table \ref{Table:Planet_parameters_Testmodels} for these fit results). Hence, all the results presented in the following sections refer to the circular-orbit model including GPs for modeling the red noise in \textit{TESS} and HARPS-N measurements. Figure \ref{Fig:Fit_ParamDistr_CornerPlot} shows the posterior distribution for the final fitted orbital parameters.

Our joint fit analysis found a linear trend in the residuals of the RV measurements after subtracting the velocity change induced by HD 20329b (see Fig. \ref{Fig:HARPSN_RVs} mid panel). This trend may be caused by another planet in the system. Based on the \textit{TESS} SAP light curves, HD 20329 appears to be a star with low photometric modulation. In addition to the low value of the $\log R'_{HK}$ activity index, HD 20329 is potentially a good candidate for long-term radial velocity monitoring to search for other planets in this system. 

We tested for any noticeable curvature in the residuals of the RV measurements. For this purpose, we added a new parameter ($\ddot{\gamma}$) to the RV modeling (circular orbit including GPs) to take this change into account. The results of the joint fit with this parameterization are presented in Table \ref{Table:Planet_parameters_curv}. From this fit, we find $\ddot{\gamma} = -0.002^{+0.004}_{-0.006}$ m/s$^3$, that is, consistent with 0, meaning that no significant curvature is found in the residuals of the fit. This was also supported by the BIC model selection criteria values, which supported an RV model with a linear trend as the best fit.

\section{Results and discussion}
\label{sec:results}
\begin{figure*}
   \centering
   \includegraphics[width=\hsize]{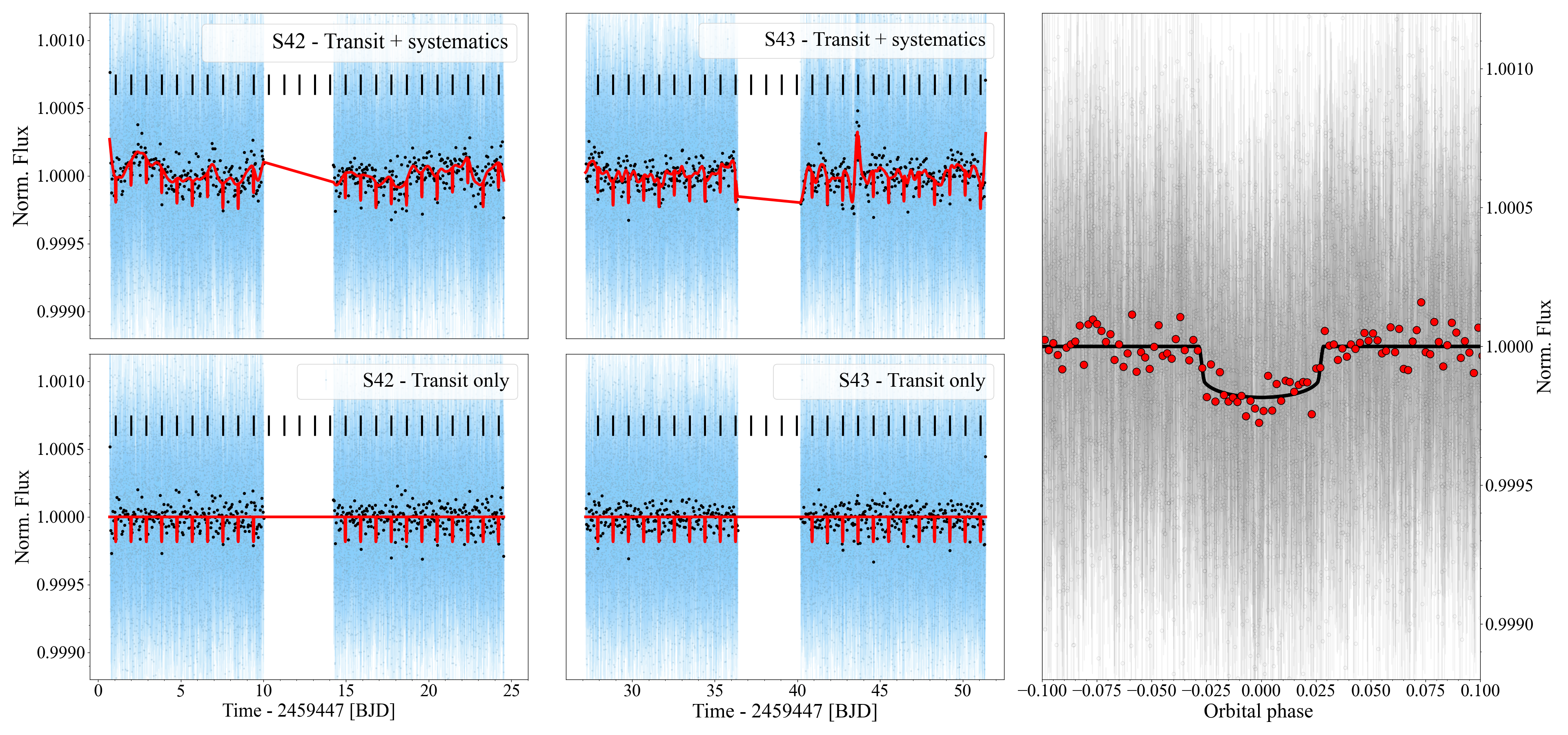}
   \caption{HD 20329 \textit{TESS} light curves and best model fit. \textit{Top and bottom left panels:} \textit{TESS} sector 42 and 43 photometry. The red line shows the best model fit with and without systematic effects, and the blue and black points are \textit{TESS} unbinned and binned data points (with a bin size of $\sim$ 1 hour). Individual transit events of HD 20329b are marked with vertical black lines. \textit{Right panel:} \textit{TESS} phase-folded light curves and best-fitting model (black line) after subtracting the photometric variability in the two \textit{TESS} sectors. The red points are binned data points.}
    \label{Fig:TESS_LightCurves}
\end{figure*}

\begin{figure*}
   \centering
   \includegraphics[width=\hsize]{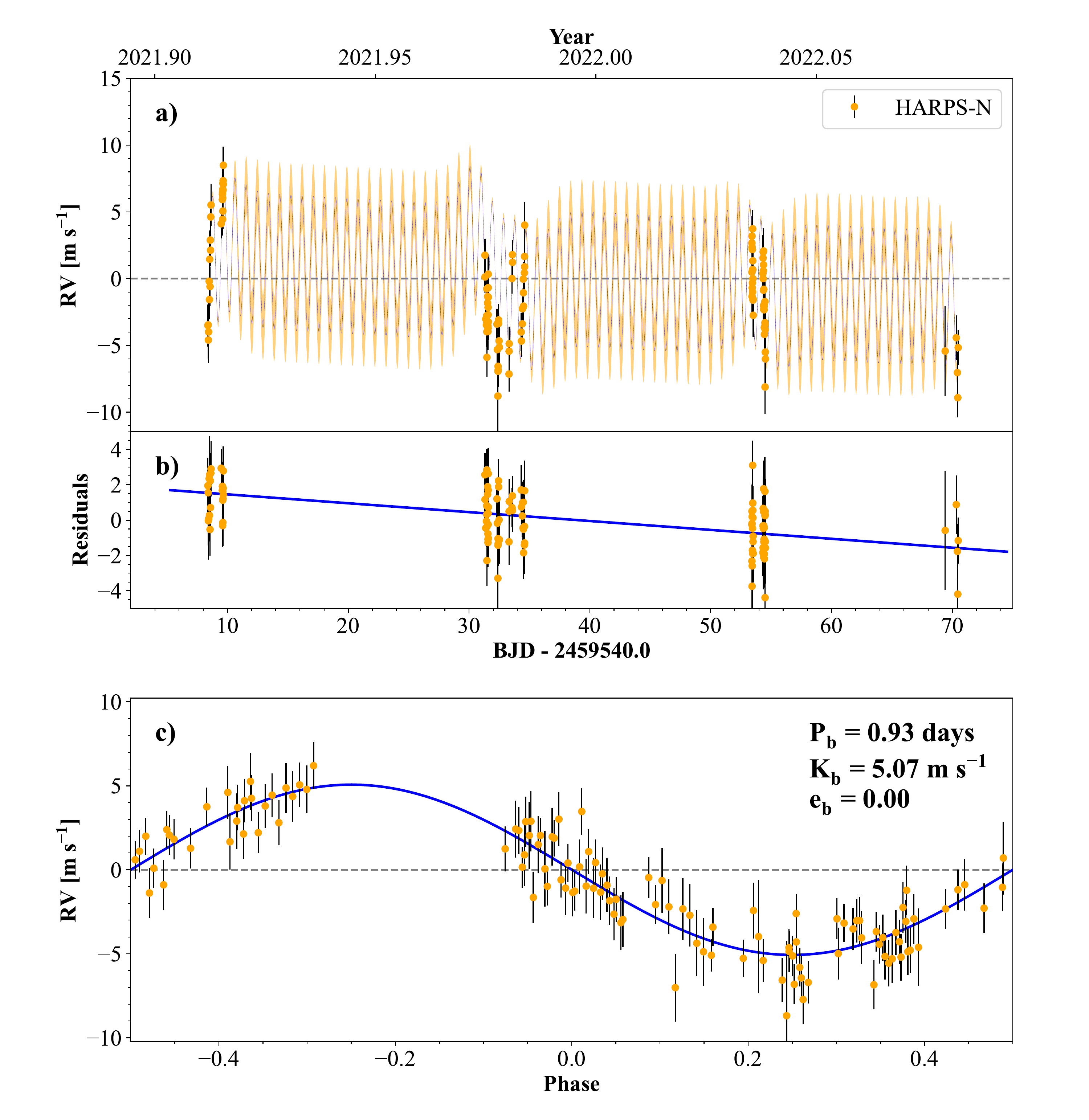}
   \caption{Radial velocity measurements of HD 20329 taken with HARPS-N. \textit{(a):} RV time series and best-fitting model (blue line) including red noise. The best model shown here was computed using the median values of the posterior distribution for each fitted parameter. The orange shaded area around the blue line represent the 1$\sigma$ uncertainty levels of the fitted model. \textit{(b):} Residuals of the fit after subtracting the single-planet Keplerian function and systematic noise component} from the RVs. \textit{(c):} Radial velocity measurements in phase after subtracting the red noise and best-fitting model (blue line).
    \label{Fig:HARPSN_RVs}
\end{figure*}

The final parameter values and 1$\sigma$ uncertainties are presented in Table \ref{Table:Planet_parameters}. Figures \ref{Fig:TESS_LightCurves} and \ref{Fig:HARPSN_RVs} show the \textit{TESS} light curves and the radial velocity measurements made by HARPS-N, respectively, and the best model found by the joint fit.

\begin{table*}
\caption{HD 20329b fitted and derived parameters, prior functions, and final values.}
\label{Table:Planet_parameters}      
\centering                          
\begin{tabular}{l c c}        
\hline\hline   
Parameter & Prior & Value \\
\hline
\multicolumn{3}{c}{Fitted orbital and transit parameters} \\
\hline
$R_{p}/R_{*}$ & $\mathcal{U}(0.005, 0.025)$ & $0.0139^{+0.0005}_{-0.0005}$ \\
$T_{c}$  [BJD] & $\mathcal{U}(2459471.7445,2459472.5445)$ & $2459472.14321^{+0.00082}_{-0.00075}$ \\
$P$ [days] & $\mathcal{U}(0.5,1.5)$ & $0.926118^{+0.000050}_{-0.000043}$ \\
$\rho_*$ [g cm$^{-3}$] & $\mathcal{N}(0.879,0.068)$ & $0.88^{+0.05}_{-0.05}$ \\
$b$ & $\mathcal{U}(0.0,1.0)$ & $0.826^{+0.017}_{-0.016}$ \\
$\gamma_0 - \langle \gamma_0 \rangle$ [m/s] & $\mathcal{U}(-6.30,9.70)$ & $3.16^{+1.53}_{-1.12}$ \\
$\dot{\gamma}$ [m/s$^{2}$] & $\mathcal{U}(-100.0, 100.0)$ & $-0.05^{+0.06}_{-0.05}$ \\
$K$ [m/s] & $\mathcal{U}(0.0,110.0)$ & $5.07^{+0.41}_{-0.42}$ \\
$\sigma_{RV}$ [m/s] & $\mathcal{U}(0.0,10.0)$ & $0.82^{+0.15}_{-0.16}$ \\
\hline
\multicolumn{3}{c}{Derived orbital parameters} \\
\hline
$e$ & & $0$ (fixed) \\ 
$a/R_*$ & & $3.42 \pm 0.06$ \\
$i$ [deg] & & $76.01 \pm 0.46$ \\
Transit duration [min] & & 75.7 \\
\hline
\multicolumn{3}{c}{Derived planet parameters} \\
\hline
R$_{p}$ [R$_{\oplus}$] & & $1.72 \pm 0.07$ \\
M$_{p}$ [M$_{\oplus}$] & & $7.42 \pm 1.09$ \\
$\rho_{p}$ [g cm$^{-3}$] & & $8.06 \pm 1.53$ \\
$g_p$ [m s$^{-2}$] & & $24.7 \pm 4.1$ \\
a [au] & & $0.0180 \pm 0.0003$ \\
T$_{eq}$ ($A_B = 0.0$) [K] & & $2141 \pm 27$ \\
T$_{eq}$ ($A_B = 0.3$) [K] & & $1958 \pm 25$ \\
$\langle F_{p} \rangle$ [10$^5$ W/m$^2$] & & $47.29 \pm 1.69$ \\
$S_{p}$ [$S_\oplus$] & & $3474 \pm 124$ \\
\hline
\multicolumn{3}{c}{Fitted LD coefficients} \\
\hline
$q_{1\;TESS}$ & $\mathcal{U}(0.0,1.0)$ & $0.32 \pm 0.02$ \\
$q_{2\;TESS}$ & $\mathcal{U}(0.0,1.0)$ & $0.37 \pm 0.02$ \\
\hline
\multicolumn{3}{c}{Derived LD coefficients} \\
\hline
$u_{1\;TESS}$ & & $0.42 \pm 0.02$ \\
$u_{2\;TESS}$ & & $0.15 \pm 0.03$ \\
\hline
\multicolumn{3}{c}{Fitted GP parameters} \\
\hline
$\log(c_1)$ TESS S42 & $\mathcal{U}(-8.0,2.3)$ & $-7.95^{+0.08}_{-0.04}$ \\
$\log(\tau_1)$ TESS S42 & $\mathcal{U}(-2.65,6.00)$ & $0.37^{+0.25}_{-0.23}$ \\
$\log(c_1)$ TESS S43 & $\mathcal{U}(-8.0,2.3)$ & $-7.98^{+0.04}_{-0.02}$ \\
$\log(\tau_1)$ TESS S43 & $\mathcal{U}(-2.65,6.00)$ & $-0.34^{+0.14}_{-0.12}$ \\
$c_2$ & $\mathcal{U}(0.0,100.0)$ & $2.41^{+4.00}_{-1.14}$ \\
$\tau_2$ & $\mathcal{U}(0.001, 150.0)$ & $1.97^{+4.99}_{-1.11}$ \\
\hline                                   
\end{tabular}
\tablefoot{$\mathcal{U}$, $\mathcal{N}$ represent uniform and normal prior functions. $A_B$ is the Bond albedo. The term $\dot{\gamma}$ was computed relative to $T_{\mathrm{base}} = 2459579.0$ BJD.}
\end{table*}

We find that HD 20329b has an orbital period of $P = 0.926118 \pm 0.00005$ days, and according to our model selection, an orbital eccentricity consistent with 0, as expected for this type of planets. Using the stellar parameter values derived from a coadded HARPS-N spectrum, we determined that HD 20329b has a radius of $R_{p} = 1.72 \pm 0.07$ R$_{\oplus}$, a mass of $M_{p} = 7.42 \pm 1.09$ M$_{\oplus}$, and a bulk density of $\rho_p = 8.06 \pm 1.53$ g cm$^{-3}$. Assuming a Bond albedo of 0.3, we find that this planet has an equilibrium temperature of $T_{eq} = 1958 \pm 25$ K.
 
\begin{figure*}
   \centering
   \includegraphics[width=\hsize]{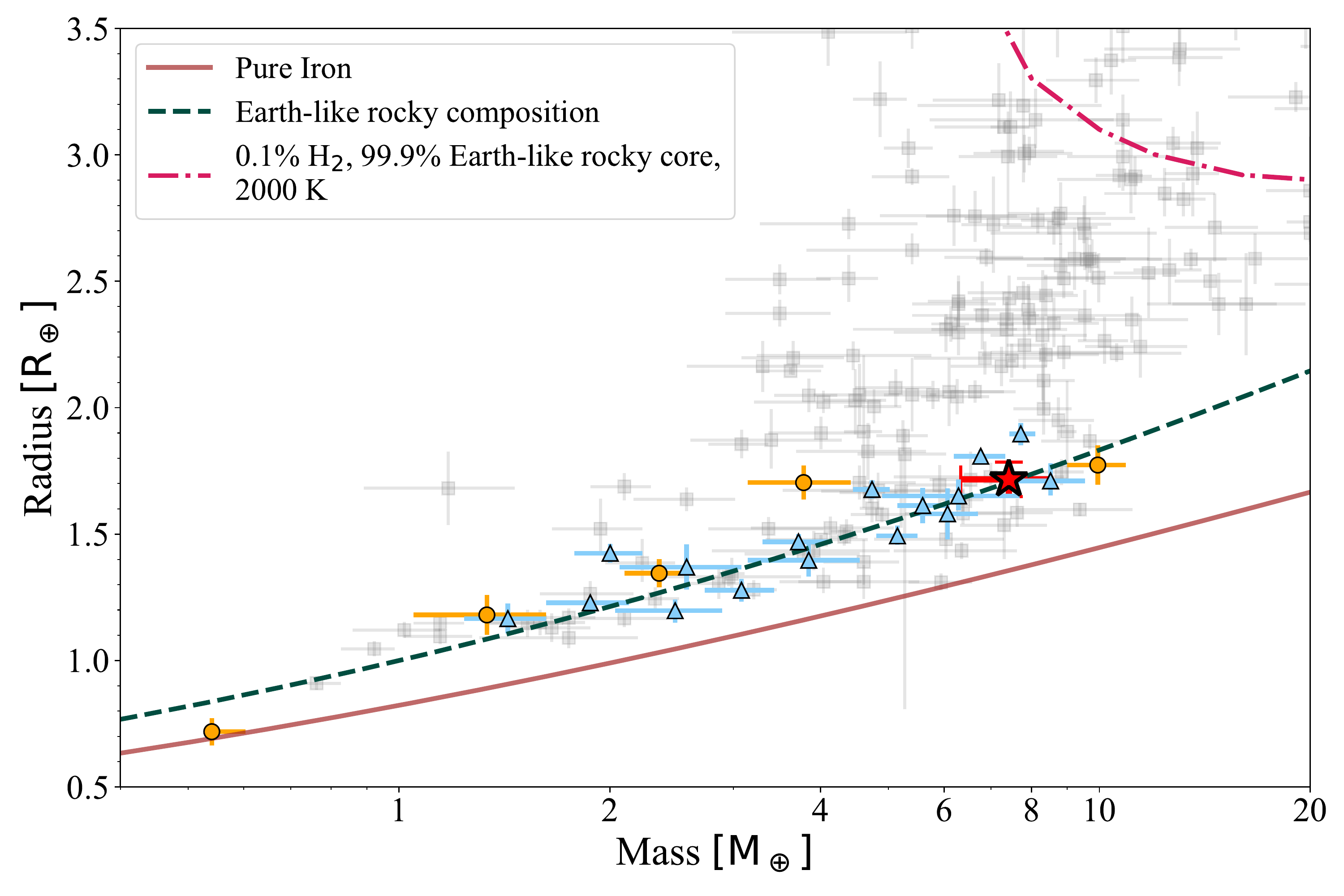}
   \caption{Mass-radius diagram for transiting planets with mass determinations with a precision better than 30\% (parameters taken from the TEPcat database; \citealp{Southworth2011}). The position of HD 20329b is shown by the red star, USP planets orbiting M-type stars ($T_{eff} \leq 4000$ K) are marked with the orange circles, USP planets around stars with $T_{eff}>4000$ K are represented by blue triangles, and non-USP planets around other types of stars are marked by gray squares. The lines in the mass-radius diagram represent the composition models of \cite{Zeng2016, Zeng2019}: planets with pure iron cores (100\% Fe, brown line), Earth-like rocky compositions (32.5\% Fe plus 67.5\% MgSiO$_3$, dashed green line), and a 99.9\% Earth-like rocky core (32.5\% Fe plus 67.5\% MgSiO$_3$) with a 0.1\% H$_2$ envelope with a temperature of 2000 K (dash-dotted pink line).}
   \label{Fig:MassRadius}
 \end{figure*}

Figure \ref{Fig:MassRadius} shows the mass versus radius distribution for known transiting planets (parameters taken from TEPcat; \citealp{Southworth2011}). USP planets around stars with $T_{eff} \leq 4000$ K and $T_{eff}>4000$ K are represented by orange circles and blue triangles, respectively. The figure also includes the rocky planet models of \cite{Zeng2016, Zeng2019} with an equilibrium temperature of 2000 K. The composition models shown in Fig. \ref{Fig:MassRadius} are planets with pure iron cores (100\% Fe), Earth-like rocky compositions (32.5\% Fe plus 67.5\% MgSiO$_3$), and Earth-like composition planet cores with 0.1\% H$_2$ gaseous envelopes. The position of HD 20329b in the mass-radius diagram agrees with other known USP planets with radii smaller than 2.0 $R_\oplus$, and it most likely indicates a rocky composition with little to no atmosphere. 

\subsection{Prospects for atmospheric characterization}
We used data from NASA Exoplanet Archive\footnote{\url{https://exoplanetarchive.ipac.caltech.edu/}} to compute the transmission spectroscopy metric (TSM) and emission spectroscopy metric (ESM) as defined by \cite{Kempton2018} for known transiting planets. These two metrics are used as indicators for the feasibility of detecting atmospheric signals during a transit (TSM) or detecting the secondary eclipse of a transiting planet (ESM) with the James Webb Space Telescope (\textit{JWST;} \citealp{Gardner2006}). We find a TSM value of $45.7$ for HD 20329b. However, \cite{Kempton2018} recommend a threshold for the TSM value higher than 90 for planets in the radius range of $1.5 < R_p < 10 \; R_\oplus$; this would mean that HD 20329b is not an ideal candidate for transmission spectroscopy follow-up if the planet holds an atmosphere. On the other hand, we find an ESM $= 10.2$. This is a favorable indicator that the secondary eclipse of this planet might be detected with \textit{JWST} given that \cite{Kempton2018} established an ESM threshold for rocky worlds of 7.5. We explore the possibility of detecting the secondary transit and phase variations of HD 20329b using \textit{TESS} data in Sect. \ref{sec:SecondaryTransit}.

\begin{figure}
   \centering
   \includegraphics[width=\hsize]{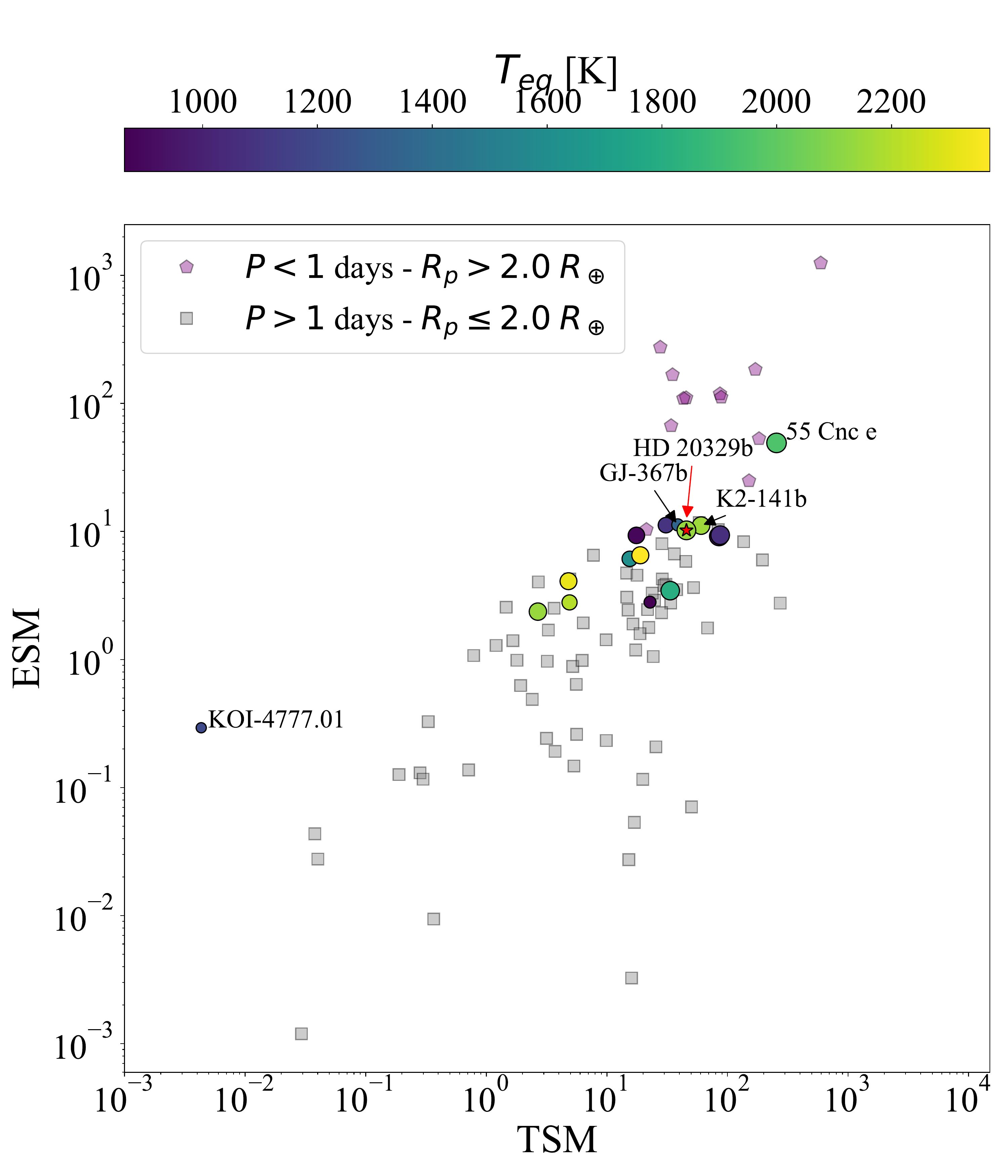}
   \caption{TSM vs ESM (\citealp{Kempton2018}) values for known transiting planets. The USP planets ($P < 1$ days) with radii smaller than 2.0 $R_\odot$ are marked by circles. The color and size of the circles represent the equilibrium temperature and radius of the planet, respectively. USPs with radii larger than 2.0 $R_\odot$ are represented with purple pentagons, and small planets ($R_p < 2.0$ $R_\odot$) with orbital periods longer than one day are marked with gray squares. The position of HD 20329b is marked by the red star and arrow. We did not label all USP planets in the plot for easy viewing.}
   \label{Fig:TSMvsESM}
 \end{figure}

Figure \ref{Fig:TSMvsESM} shows the TSM and ESM values for known transiting planets. We focused on USP period planets and planets with radii smaller than 2 $R_\oplus$. The planets with the most similar TSM and ESM values to HD 20329b are GJ 367b and K2-141b. GJ 367b was discovered by \cite{Lam2021}. The planet is a sub-Earth that orbits an M dwarf with an orbital period of $P = 0.321$ days (7.7 hours) and has a mass and radius of $0.546 \pm 0.078$ $M_\oplus$ and $0.718 \pm 0.054$ $R_\oplus$. The mean bulk density of this planet is similar to that of HD 20329b with $\rho_p = 8.1 \pm 2.2$ g cm$^{-3}$. Because GJ 367b orbits an M dwarf, this planet receives less flux than HD 20329b, and their equilibrium temperatures differ by $\sim 360$ K ($T_{eq} = 1597 \pm 39$ K for GJ 367b assuming $A_B = 0.3$). K2-141b was discovered independently by \cite{Malavolta2018} and \cite{Barragan2018} using data from the \textit{Kepler} extended mission \textit{K2} (\citealp{Howell2014}). This planet is a rocky super-Earth orbiting a K dwarf with a period of $P = 0.280$ days ($\sim 6.7$ hours), and with a mass and radius of $5.08 \pm 0.041$ $M_\oplus$ and $1.51 \pm 0.05$ $R_\oplus$. K2-141b has a similar mean bulk density ($\rho_p = 8.1 \pm 1.1$ g cm$^{-3}$) and equilibrium temperature ($T_{eq} = 2039^{+87}_{-48}$ K, \citealp{Barragan2018}) as HD 20329b.
 
\subsection{Secondary eclipse in \textit{TESS} data}
\label{sec:SecondaryTransit}

In this section we describe our efforts to explore the detection of the secondary transit and phase variations of HD 20329b using \textit{TESS} data. We modeled the \tess light curves from sectors 42 and 43 using the \texttt{PhaseCurveLPF} phase-curve model implemented in \pytransit (see \citealp{Parviainen2021}). The phase-curve model
includes the effects from thermal emission, reflection, ellipsoidal variation, and Doppler boosting, but we forced the ellipsoidal variation and Doppler boosting
amplitudes to zero with narrow normal priors because the planet-to-star mass ratio is too low for these two effects to have any practical importance. Furthermore,
we set the geometric albedo to zero (again, using a narrow normal prior) because the reflection and emission components are strongly degenerate, and we are primarily
interested in determining whether an eclipse signal exists in the light curves.

With these exceptions, the phase curve model includes 12 free parameters (transit center, orbital period, stellar density, impact parameter, $\sqrt{e} \cos{\omega}$,
$\sqrt{e}\sin{\omega}$, planet-star area ratio, log$_{10}$ dayside planet-star flux ratio from emission, log$_{10}$ nightside planet-star flux ratio from emission,
emission peak offset, and two limb-darkening coefficients). We set weakly informative normal priors on the transit center and orbital 
period based on the information in ExoFOP, and an informative normal prior, $N(0.88, 0.07)$, on the stellar density based on the stellar
characterisation described in Sect.~\ref{sec:stellar_param}. We also forced a circular orbit by setting tight zero-centered normal
priors on $\sqrt{e} \cos{\omega}$ and $\sqrt{e}\sin{\omega}$ and constrained the limb-darkening coefficients using priors calculated with \ldtk.
The baseline flux variations were modeled as a time-dependent GP using \celerite (\citealp{ForemanMackey2017}). Each \textit{TESS} sector was assigned a separate GP with its own hyperparameters, and the hyperparameters were kept free and marginalized over during the posterior sampling. The day- and nightside log$_{10}$ flux ratios had uniform priors from -3 to 0.

The phase-curve model parameter posteriors were obtained as usual. We first determined the global posterior maximum using a global optimization method, and then 
sampled the posterior using MCMC sampling. 

\begin{figure*}
   \centering
   \includegraphics[width=\textwidth]{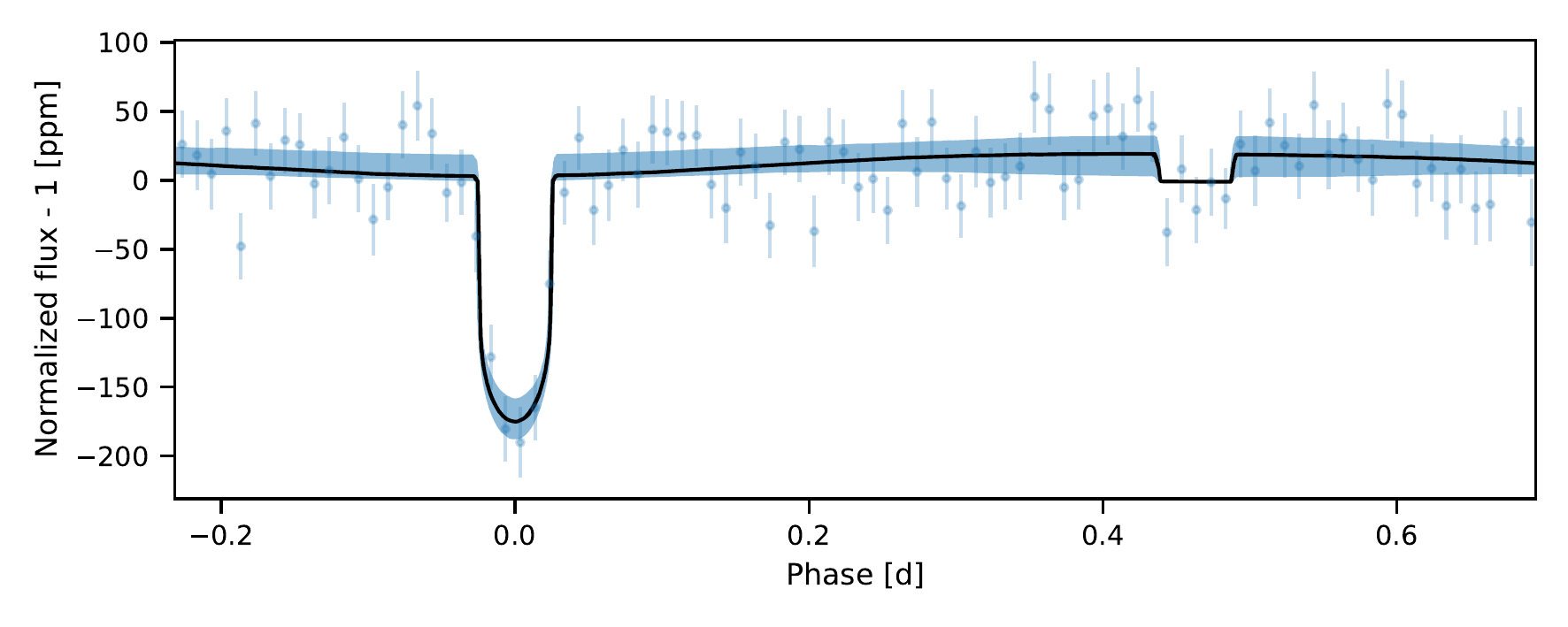}
   \caption{\tess observations from sectors 42 and 43 folded over the orbital phase (in days) and binned (for visualization) with the median posterior phase-curve model (black line) and its 16th and 84th percentile limits (blue shading). The eclipse is modeled assuming a uniform stellar disk with a depth scaled from 0 (eclipse) to 1 (out of eclipse).}
   \label{fig:phasecurve}
 \end{figure*}

\begin{figure}
   \centering
   \includegraphics[width=\columnwidth]{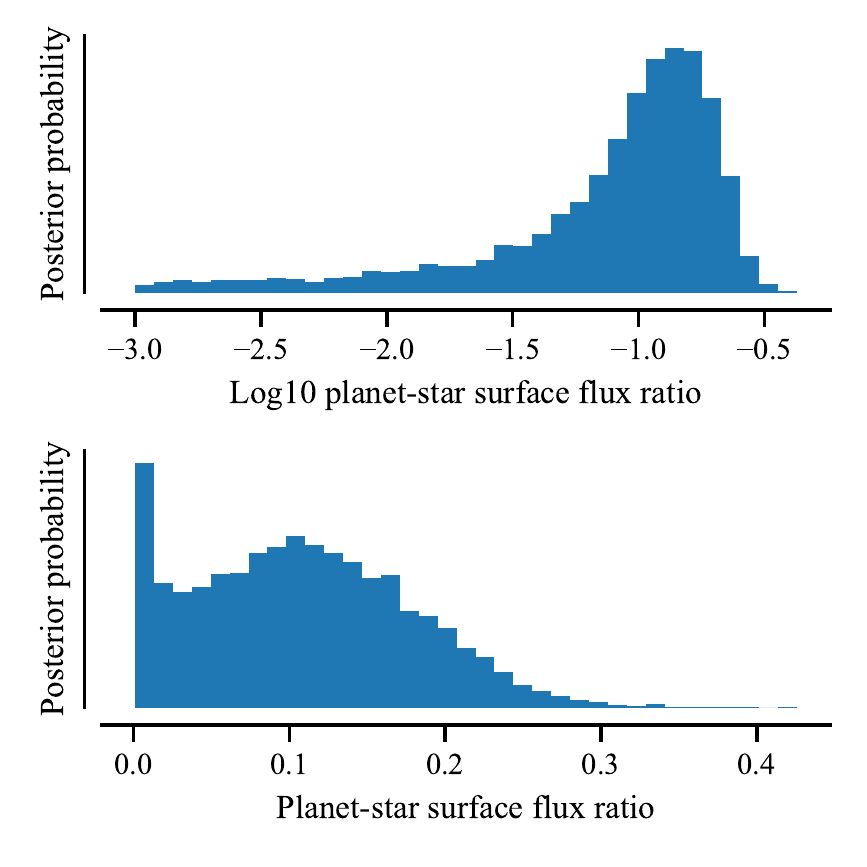}
   \caption{Posterior probability densities for the dayside log$_{10}$ planet-to-star flux ratio used as a sampling parameter and the flux ratio.}
   \label{fig:phase_posteriors}
 \end{figure}

The results from the phase-curve modeling (Fig.~\ref{fig:phasecurve}) support the existence of an eclipse signal in the \tess data. 
The log$_{10}$ dayside flux ratio posterior has its median at -0.98, and the dayside flux ratio estimate\footnote{The estimates are based on the posterior median and 16th and 84th posterior percentiles.} is 
$11\%\pm8\%$, which corresponds to an eclipse depth of $20\pm14$~ppm, as shown in Fig.~\ref{fig:phase_posteriors}. The nightside
flux ratio is not constrained. The dayside posterior
does not exclude a no-eclipse scenario (i.e., with a planet-star flux ratio of zero), but does support a relatively significant planetary emission signal. We used
log$_{10}$ flux ratio as a sampling parameter, and a uniform prior on it sets a reciprocal (log-uniform) prior on the flux ratio itself. We would expect a
uniform log$_{10}$ flux ratio posterior distribution in the absence of an eclipse signal, which would allow us to give only an upper limit for the flux ratio. However,
the log$_{10}$ dayside flux ratio posterior has a strong mode with a tail toward lower values, which indicates that models with little to no eclipse signal can also explain the observations. 

Curious about the possible eclipse detection, we mapped the brightness temperatures and geometric albedos that might explain the dayside flux ratio posterior.
We modeled the flux ratio as the sum of emission and reflection components. The emission component was calculated using the BT-Settl spectra\footnote{The BT-Settl
spectra are evaluated using an interpolator created with \texttt{pytransit.stars.create\char`_bt\char`_settl\char`_interpolator}.} , where we assumed the host star to have an effective
temperature of 5640~K and kept the planet temperature as a free parameter. The reflection component had the geometric albedo as a free parameter, and the remaining parameters (semi-major axis and planet-star radius ratio) were fixed to the posterior median values from the phase-curve modeling. 

\begin{figure}
   \centering
   \includegraphics[width=\columnwidth]{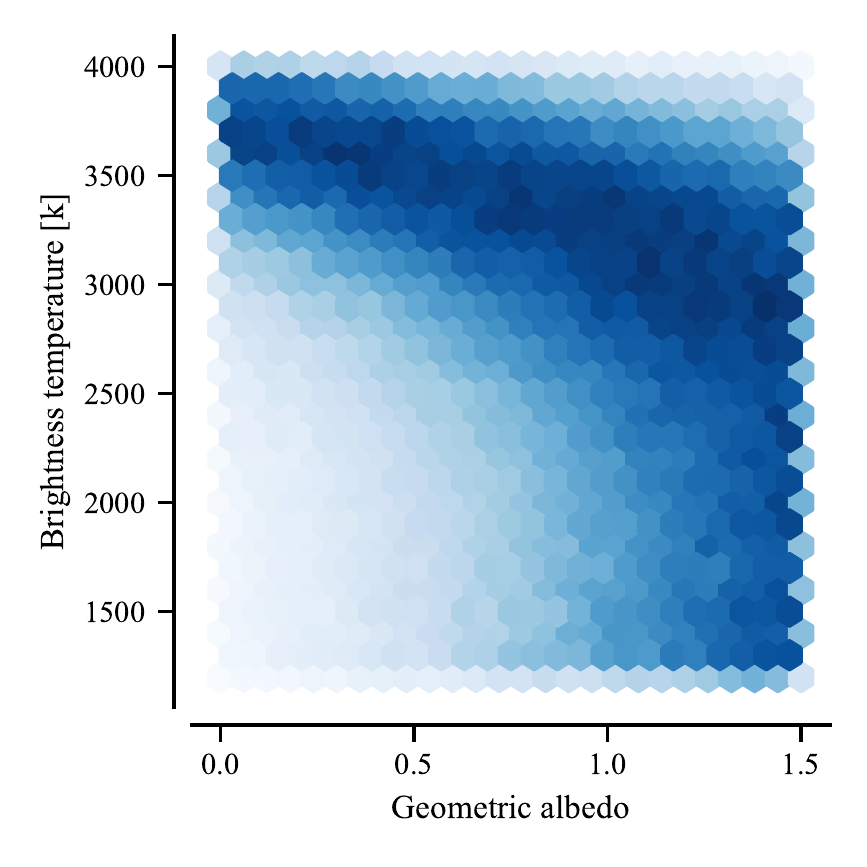}
   \caption{Posterior probability densities for the geometric albedo and brightness temperature.}
   \label{fig:phase_albedo_and_temperature}
 \end{figure}

We estimated the brightness temperature and geometric albedo posterior via MCMC sampling using a Gaussian kernel density estimate (kde) of the log$_{10}$ dayside flux ratio posterior as a target distribution (i.e., the log likelihood is based on the kde log pdf evaluated for the log$_{10}$ flux ratio model values). The final
geometric albedo versus brightness distribution is shown in Fig.~\ref{fig:phase_albedo_and_temperature}. The dayside flux ratio posterior from the phase-curve modeling suggests a very high brightness temperature for the planet that can be reduced only by high geometric albedos. The median posterior brightness temperature for $A_\mathrm{g} < 0.25$ is $\approx3500$~K, while for $0.25 < A_\mathrm{g} < 0.5,$ it is $\approx3300$~K. However, the median posterior value does not characterize the marginalized brightness temperature posterior well because it changes from a distribution with a clear mode closer to a uniform distribution as the geometric albedo increases. The upper limit for the brightness temperature decreases with increasing geometric albedo (as it should), but the 99th posterior percentile is $\sim4000$~K over the $A_\mathrm{g}$ range from 0 to 1. To support this conclusion, we also modeled the light curve with the transit and light-curve modeler code (\citealp{csizmadia2020}), which obtained similar results (see Appendix~\ref{Table:Planet_parameters_Testmodels_TLCM}).

\subsection{Additional planets in the system}
As pointed out in \cite{SanchisOjeda2014} and \cite{Winn2018}, USP planets are often found in multiple-planet systems. We searched for additional transit events in \textit{TESS} data using \texttt{Transit Least Squares} (\texttt{TLS}; \citealp{Hippke2019}). We used the results of the joint model fit to mask the transit events of planet b and to remove the photometric noise in the \textit{TESS} time series modeled by the GPs. We also tested median filters with different window sizes to remove the photometric variability in the light curves and ran \texttt{TLS} multiple times. We found no evidence of other significant transit events in \textit{TESS} data. The transits that \texttt{TLS} detected are likely spurious signals, with at least one transit occurring at the beginning or end of a block of \textit{TESS} observations where systematic effects due to the pointing of the satellite are stronger. In addition, the detected events presented a low signal detection efficiency (SDE) in comparison to HD 20329b, which \texttt{TLS} detected.

The time baseline of our RV measurements is 62 days because we do not detect a significant curvature in the RV residuals (see end of Sect. \ref{sec:transit_and_rv_fit}). This means that the object that causes the RV slope must have an orbital period longer than at least four times the baseline. Using Kepler's third law and assuming $M_\star \gg M_p$, an object with $P = 120$ days has an orbital semi-major axis of $0.46$ AU. Based on the maximum $\Delta RV$ difference of the residuals and setting this value as the semi-amplitude of an object with orbital period $P = 120$ days, we find a minimum mass of $M_p \sin(i) = 0.17$ M$_{\mathrm{Jup}}$. 

If the radial velocity change seen in the RV residuals is caused by a stellar mass object, this might affect the projected motion of the star in the sky. \cite{Brandt2021} used the \textit{Hipparcos} (\citealp{ESA1997}) and \textit{Gaia} EDR3 (\citealp{Lindegren2021}) data to create a catalog of proper motions with a time baseline of $\sim$24 years (1991.25, 2015.5) to identify astrometrically accelerating stars. For HD 20329, the \textit{Gaia} proper motions are $\vec{\mu}_G = (111.77 \pm 0.030, -202.41 \pm 0.025)$ mas yr$^{-1}$ and the derived \textit{Hipparcos}-\textit{Gaia} proper motions are $\vec{\mu}_{HG} = (111.73 \pm 0.035, -202.43 \pm 0.026)$ mas yr$^{-1}$, meaning that the star presented a change in proper motion of $\Delta \mu = |\vec{\mu}_G - \vec{\mu}_G| = 0.047 \pm 0.044 $ mas yr $^{-1}$ over 24 years. This null acceleration is consistent with the \textit{Gaia} EDR3 assessment discussed in section \ref{sec:HighResImg}, in which \textit{Gaia} measurements favor a single-star model. The slope in the RV residuals is therefore likely caused by a substellar object. 
 
\section{Conclusions}
\label{sec:conclusions}
We reported the discovery of HD 20329b, an ultra-short-period transiting planet around a solar-type star. Observations made by the NASA \textit{TESS} mission led to the initial detection of the transits. Follow-up radial velocity measurements taken with the HARPS-N spectrograph made it possible to confirm the planetary nature of the transiting object and to establish its mass and radius. 

From a coadded HARPS-N mean stellar spectrum, we estimate that the host star has an effective temperature of $\mathrm{T}_\mathrm{eff}=5596 \pm 50 $ K, a metallicity of $[\mathrm{Fe}/\mathrm{H}] = -0.07 \pm 0.06$ dex, and a derived stellar mass and radius of $\mathrm{M}_\star = 0.90\pm 0.05$ $\mathrm{M}_\odot$ and $\mathrm{R}_\star = 1.13\pm 0.02$ $\mathrm{R}_\odot$.

We analyzed \textit{TESS} photometric data from sectors 42 and 43 and radial velocity measurements taken with the HARPS-N spectrograph. The transit observations made by \textit{TESS} and the radial velocity measurements were fit simultaneously using an MCMC procedure that included Gaussian processes to model the systematic effects present in the light curves and RV measurements. We find that HD 20329b has a radius of $\mathrm{R}_\mathrm{p} = 1.72 \pm 0.07$ $\mathrm{R}_\oplus$ and a mass of $\mathrm{M}_\mathrm{p} = 7.42 \pm 1.09$ $\mathrm{M}_\oplus$, and that it orbits its star with a period of $0.926118 \pm 0.50\times 10^{-4}$ days. We derived a mean bulk density of $\rho_\mathrm{p} = 8.06 \pm 1.53$ g cm$^{-3}$, indicating a likely rocky composition. We note that after subtracting the planetary signal of the USP planet from the RV measurements, the RV residuals present a slope that might indicate the presence of an additional low-mass object orbiting HD 20329. 

The ESM for HD 20329b indicates that this planet presents favorable conditions for secondary transit follow-up with the \textit{JWST}. We used a simple phase-curve model including reflected light and thermal emission to search for the secondary transit and phase variations in \textit{TESS} light curves. Our results support the existence of a significant, although not conclusive, eclipse signal in the \tess data, with a dayside flux ratio of 11\% and a relatively strong planetary emission signal. Our modeling indicates a brightness temperature of $\sim$3500 K for low geometric albedo values  ($A_\mathrm{g} < 0.25$) and an upper limit on the brightness temperature of $\sim$4000 K over the range $A_\mathrm{g} \in [0,1]$. Precise observations, preferable in the IR, are needed to confirm these results.

Overall, HD 20329b is a new addition to the $\sim$120 currently known USP planets. It presents very favorable metrics for a secondary transit detection with the \textit{JWST} and for radial velocity follow-up to search for additional planets in the system.


\begin{acknowledgements}
TM acknowledges financial support from the Spanish Ministry
of Science and Innovation (MICINN) through the Spanish State
Research Agency, under the Severo Ochoa Program 2020-2023
(CEX2019-000920-S) as well as support from the ACIISI, Consejer\'{i}a de Econom\'{i}a, Conocimiento y Empleo del Gobierno
de Canarias and the European Regional Development Fund (ERDF) under grant with reference  PROID2021010128. R.L. acknowledges funding from University of La Laguna through the Margarita Salas Fellowship from the Spanish Ministry of Universities ref. UNI/551/2021-May 26 under the EU Next Generation funds and financial support from the Spanish Ministerio de Ciencia e Innovaci\'on, through project PID2019-109522GB-C52, and the Centre of Excellence "Severo Ochoa" award to the Instituto de Astrofísica de Andalucía (SEV-2017-0709). PK acknolwedges the support from grant LTT-20015. K.W.F.L. was supported by Deutsche Forschungsgemeinschaft grants RA714/14-1 within the DFG Schwerpunkt SPP 1992, Exploring the Diversity of Extrasolar Planets. HJD acknowledges support from the Spanish Research Agency of the Ministry of Science and Innovation (AEI-MICINN) under the grant PID2019-107061GB-C66, DOI: 10.13039/501100011033. C.M.P. and J.K. gratefully acknowledge the support of the Swedish National Space Agency (DNR 65/19 2020-00104). This paper includes data collected by the \textit{TESS} mission, which are publicly available from the Mikulski Archive for Space Telescopes (MAST). Funding for the \textit{TESS} mission is provided by NASA's Science Mission directorate. Resources supporting this work were provided by the NASA High-End Computing (HEC) Program through the NASA Advanced Supercomputing (NAS) Division at Ames Research Center for the production of the SPOC data products. We acknowledge the use of public \textit{TESS} data from pipelines at the \textit{TESS} Science Office and at the \textit{TESS} Science Processing Operations Center. This research has made use of the Exoplanet Follow-up Observation Program website, which is operated by the California Institute of Technology, under contract with the National Aeronautics and Space Administration under the Exoplanet Exploration Program. This publication makes use of data products from the AAVSO Photometric All Sky Survey (APASS). Funded by the Robert Martin Ayers Sciences Fund and the National Science Foundation. This work has made use of data from the European Space Agency (ESA) mission
{\it Gaia} (\url{https://www.cosmos.esa.int/gaia}), processed by the {\it Gaia}
Data Processing and Analysis Consortium (DPAC,
\url{https://www.cosmos.esa.int/web/gaia/dpac/consortium}). Funding for the DPAC
has been provided by national institutions, in particular the institutions
participating in the {\it Gaia} Multilateral Agreement. This work made use of \texttt{tpfplotter} by J. Lillo-Box (publicly available in \url{www.github.com/jlillo/tpfplotter}), which also made use of the python packages \texttt{astropy}, \texttt{lightkurve}, \texttt{matplotlib} and \texttt{numpy}. This work is partly supported by JSPS KAKENHI Grant Numbers
JP17H04574, JP18H05442, JST CREST Grant Number JPMJCR1761, and the
Astrobiology Center of National Institutes of Natural Sciences (NINS)
(Grant Number AB031010).
This paper is based on observations made with the MuSCAT2 instrument,
developed by ABC, at Telescopio Carlos S\'anchez operated on the island
of Tenerife by the IAC in the Spanish Observatorio del Teide.
\end{acknowledgements}

%
%

\bibliographystyle{aa}
\bibliography{references}

\begin{appendix}

\section{Radial velocity data}
Radial velocity and activity indices measurements are available electronically at CDS.

\null\newpage
\clearpage
\onecolumn
\begin{tiny}
\begin{center}
\tablefirsthead{
\hline
\hline
\multicolumn{1}{r}{BJD$_\mathrm{TBD}$} &
\multicolumn{1}{r}{RV} &
\multicolumn{1}{r}{$\sigma_\mathrm{RV}$} &
\multicolumn{1}{r}{BIS} &
\multicolumn{1}{r}{$\sigma_\mathrm{BIS}$} &
\multicolumn{1}{r}{CCF\_FWHM} &
\multicolumn{1}{r}{CCF\_CTR} &
\multicolumn{1}{r}{$\mathrm{\log{R^{`}_{HK}}}$} &
\multicolumn{1}{r}{$\mathrm{\sigma_{\log{R^{`}_{HK}}}}$} &
\multicolumn{1}{r}{SNR} &
\multicolumn{1}{r}{$\mathrm{T_{exp}}$}\\ 
\multicolumn{1}{r}{-2457000} &
\multicolumn{1}{r}{($\mathrm{m\,s^{-1}}$)} &
\multicolumn{1}{r}{($\mathrm{m\,s^{-1}}$)} &
\multicolumn{1}{r}{($\mathrm{m\,s^{-1}}$)} &
\multicolumn{1}{r}{($\mathrm{m\,s^{-1}}$)} &
\multicolumn{1}{r}{($\mathrm{km\,s^{-1}}$)} &
\multicolumn{1}{r}{(\%)} &
\multicolumn{1}{r}{---} &
\multicolumn{1}{r}{---} &
\multicolumn{1}{r}{@550nm} &
\multicolumn{1}{r}{(s)}\\ 
\hline
}
\tablehead{
\multicolumn{11}{c}
{{\bfseries \tablename\ \thetable{}} -- continued from previous page.}\\
\hline
\hline
\multicolumn{1}{r}{BJD$_\mathrm{TBD}$} &
\multicolumn{1}{r}{RV} &
\multicolumn{1}{r}{$\sigma_\mathrm{RV}$} &
\multicolumn{1}{r}{BIS} &
\multicolumn{1}{r}{$\sigma_\mathrm{BIS}$} &
\multicolumn{1}{r}{CCF\_FWHM} &
\multicolumn{1}{r}{CCF\_CTR} &
\multicolumn{1}{r}{$\mathrm{\log{R^{`}_{HK}}}$} &
\multicolumn{1}{r}{$\mathrm{\sigma_{\log{R^{`}_{HK}}}}$} &
\multicolumn{1}{r}{SNR} &
\multicolumn{1}{r}{$\mathrm{T_{exp}}$}\\ 
\multicolumn{1}{r}{-2457000} &
\multicolumn{1}{r}{($\mathrm{m\,s^{-1}}$)} &
\multicolumn{1}{r}{($\mathrm{m\,s^{-1}}$)} &
\multicolumn{1}{r}{($\mathrm{m\,s^{-1}}$)} &
\multicolumn{1}{r}{($\mathrm{m\,s^{-1}}$)} &
\multicolumn{1}{r}{($\mathrm{km\,s^{-1}}$)} &
\multicolumn{1}{r}{(\%)} &
\multicolumn{1}{r}{---} &
\multicolumn{1}{r}{---} &
\multicolumn{1}{r}{@550nm} &
\multicolumn{1}{r}{(s)}\\ 
\hline
}
\tabletail{
\hline
\multicolumn{11}{c}{{Continued on next page}}\\
\hline
}
\tablelasttail{
\hline
}
\tablecaption{Radial velocities and spectral activity indicators measured from TNG/HARPS-N spectra measured with DRS.\label{table-TOI-4524-tng_harpn-0120-drs-complete_output}}
\begin{supertabular}{rrrrrrrrrrr}
     2548.38925 &      -71541.952 &           1.406 &         -33.408 &           1.988 &           6.706 &          49.557 &         -5.0304 &          0.0240 &            64.2 &          1800.0\\
     2548.41148 &      -71540.799 &           1.113 &         -31.256 &           1.574 &           6.707 &          49.600 &         -5.0235 &          0.0160 &            78.5 &          1800.0\\
     2548.43064 &      -71541.531 &           1.198 &         -28.501 &           1.694 &           6.710 &          49.567 &         -5.0372 &          0.0184 &            73.5 &          1800.0\\
     2548.44897 &      -71549.223 &           2.426 &         -28.041 &           3.431 &           6.715 &          49.397 &         -4.9776 &          0.0463 &            40.8 &          1297.7\\
     2548.49747 &      -71538.122 &           1.524 &         -32.264 &           2.156 &           6.714 &          49.672 &         -5.0946 &          0.0305 &            59.6 &          1800.0\\
     2548.51778 &      -71537.561 &           1.403 &         -28.272 &           1.984 &           6.712 &          49.672 &         -5.1202 &          0.0283 &            63.9 &          1800.0\\
     2548.53804 &      -71536.048 &           2.311 &         -18.694 &           3.269 &           6.707 &          49.770 &         -5.1484 &          0.0645 &            42.2 &          1800.0\\
     2548.56765 &      -71536.879 &           1.561 &         -29.939 &           2.208 &           6.707 &          49.745 &         -5.0651 &          0.0300 &            58.3 &          1800.0\\
\end{supertabular}
\end{center}
\end{tiny}

\null\newpage
\clearpage
\onecolumn
\begin{tiny}
\begin{center}
\tablefirsthead{
\hline
\hline
\multicolumn{1}{r}{BJD$_\mathrm{TBD}$} &
\multicolumn{1}{r}{RV} &
\multicolumn{1}{r}{$\sigma_\mathrm{RV}$} &
\multicolumn{1}{r}{CRX} &
\multicolumn{1}{r}{$\sigma_\mathrm{CRX}$} &
\multicolumn{1}{r}{dlW} &
\multicolumn{1}{r}{$\sigma_\mathrm{dlW}$} &
\multicolumn{1}{r}{$\mathrm{H_{\alpha}}$} &
\multicolumn{1}{r}{$\mathrm{\sigma_{H_{\alpha}}}$} &
\multicolumn{1}{r}{$\mathrm{NaD_{1}}$} &
\multicolumn{1}{r}{$\mathrm{\sigma_{NaD_{1}}}$} &
\multicolumn{1}{r}{$\mathrm{NaD_{2}}$} &
\multicolumn{1}{r}{$\mathrm{\sigma_{NaD_{2}}}$}\\ 
\multicolumn{1}{r}{-2457000} &
\multicolumn{1}{r}{($\mathrm{m\,s^{-1}}$)} &
\multicolumn{1}{r}{($\mathrm{m\,s^{-1}}$)} &
\multicolumn{1}{r}{($\mathrm{m\,s^{-1}\,Np^{-1}}$)} &
\multicolumn{1}{r}{($\mathrm{m\,s^{-1}\,Np^{-1}}$)} &
\multicolumn{1}{r}{($\mathrm{m^2\,s^{-2}}$)} &
\multicolumn{1}{r}{($\mathrm{m^2\,s^{-2}}$)} &
\multicolumn{1}{r}{---} &
\multicolumn{1}{r}{---} &
\multicolumn{1}{r}{---} &
\multicolumn{1}{r}{---} &
\multicolumn{1}{r}{---} &
\multicolumn{1}{r}{---}\\
\hline
}
\tablehead{
\multicolumn{13}{c}
{{\bfseries \tablename\ \thetable{}} -- continued from previous page.}\\
\hline
\hline
\multicolumn{1}{r}{BJD$_\mathrm{TBD}$} &
\multicolumn{1}{r}{RV} &
\multicolumn{1}{r}{$\sigma_\mathrm{RV}$} &
\multicolumn{1}{r}{CRX} &
\multicolumn{1}{r}{$\sigma_\mathrm{CRX}$} &
\multicolumn{1}{r}{dlW} &
\multicolumn{1}{r}{$\sigma_\mathrm{dlW}$} &
\multicolumn{1}{r}{$\mathrm{H_{\alpha}}$} &
\multicolumn{1}{r}{$\mathrm{\sigma_{H_{\alpha}}}$} &
\multicolumn{1}{r}{$\mathrm{NaD_{1}}$} &
\multicolumn{1}{r}{$\mathrm{\sigma_{NaD_{1}}}$} &
\multicolumn{1}{r}{$\mathrm{NaD_{2}}$} &
\multicolumn{1}{r}{$\mathrm{\sigma_{NaD_{2}}}$}\\ 
\multicolumn{1}{r}{-2457000} &
\multicolumn{1}{r}{($\mathrm{m\,s^{-1}}$)} &
\multicolumn{1}{r}{($\mathrm{m\,s^{-1}}$)} &
\multicolumn{1}{r}{($\mathrm{m\,s^{-1}\,Np^{-1}}$)} &
\multicolumn{1}{r}{($\mathrm{m\,s^{-1}\,Np^{-1}}$)} &
\multicolumn{1}{r}{($\mathrm{m^2\,s^{-2}}$)} &
\multicolumn{1}{r}{($\mathrm{m^2\,s^{-2}}$)} &
\multicolumn{1}{r}{---} &
\multicolumn{1}{r}{---} &
\multicolumn{1}{r}{---} &
\multicolumn{1}{r}{---} &
\multicolumn{1}{r}{---} &
\multicolumn{1}{r}{---}\\ 
\hline
}
\tabletail{
\hline
\multicolumn{13}{c}{{Continued on next page}}\\
\hline
}
\tablelasttail{
\hline
}
\tablecaption{Radial velocities and spectral activity indicators measured from TNG/HARPS-N spectra measured with SERVAL.\label{table-TOI-4524-tng_harpn-0120-srv-complete_output}}
\begin{supertabular}{rrrrrrrrrrrrr}
     2548.38925 &          -0.345 &           1.351 &         -19.144 &          11.601 &          -4.199 &           2.981 &          0.4387 &          0.0013 &          0.3088 &          0.0014 &          0.4224 &          0.0018\\
     2548.41148 &          -0.304 &           1.099 &         -12.633 &           9.602 &          -6.084 &           2.228 &          0.4396 &          0.0011 &          0.3084 &          0.0012 &          0.4207 &          0.0015\\
     2548.43064 &          -1.438 &           1.128 &          -3.862 &           9.985 &          -4.775 &           2.483 &          0.4373 &          0.0012 &          0.3068 &          0.0013 &          0.4171 &          0.0016\\
     2548.44897 &          -0.831 &           2.162 &           6.980 &          19.018 &           5.030 &           3.987 &          0.4401 &          0.0022 &          0.3165 &          0.0024 &          0.4216 &          0.0029\\
     2548.49747 &           2.964 &           1.305 &           5.994 &          11.513 &         -13.729 &           3.001 &          0.4360 &          0.0014 &          0.3171 &          0.0016 &          0.4224 &          0.0020\\
     2548.51778 &           1.592 &           1.236 &          -3.673 &          10.979 &         -14.972 &           2.273 &          0.4374 &          0.0014 &          0.3175 &          0.0015 &          0.4218 &          0.0018\\
     2548.53804 &           4.609 &           2.015 &          14.245 &          17.746 &         -27.117 &           5.311 &          0.4461 &          0.0021 &          0.3180 &          0.0023 &          0.4217 &          0.0028\\
     2548.56765 &           2.566 &           1.229 &         -15.540 &          10.624 &         -22.572 &           3.374 &          0.4425 &          0.0014 &          0.3185 &          0.0016 &          0.4236 &          0.0020\\
\end{supertabular}
\end{center}
\end{tiny}
 %

\twocolumn
\section{Stellar rotation from ASAS-SN photometry}
\label{Appendix:StellarRot}
\begin{figure*}[htbp]
   \centering
   \includegraphics[width=\hsize]{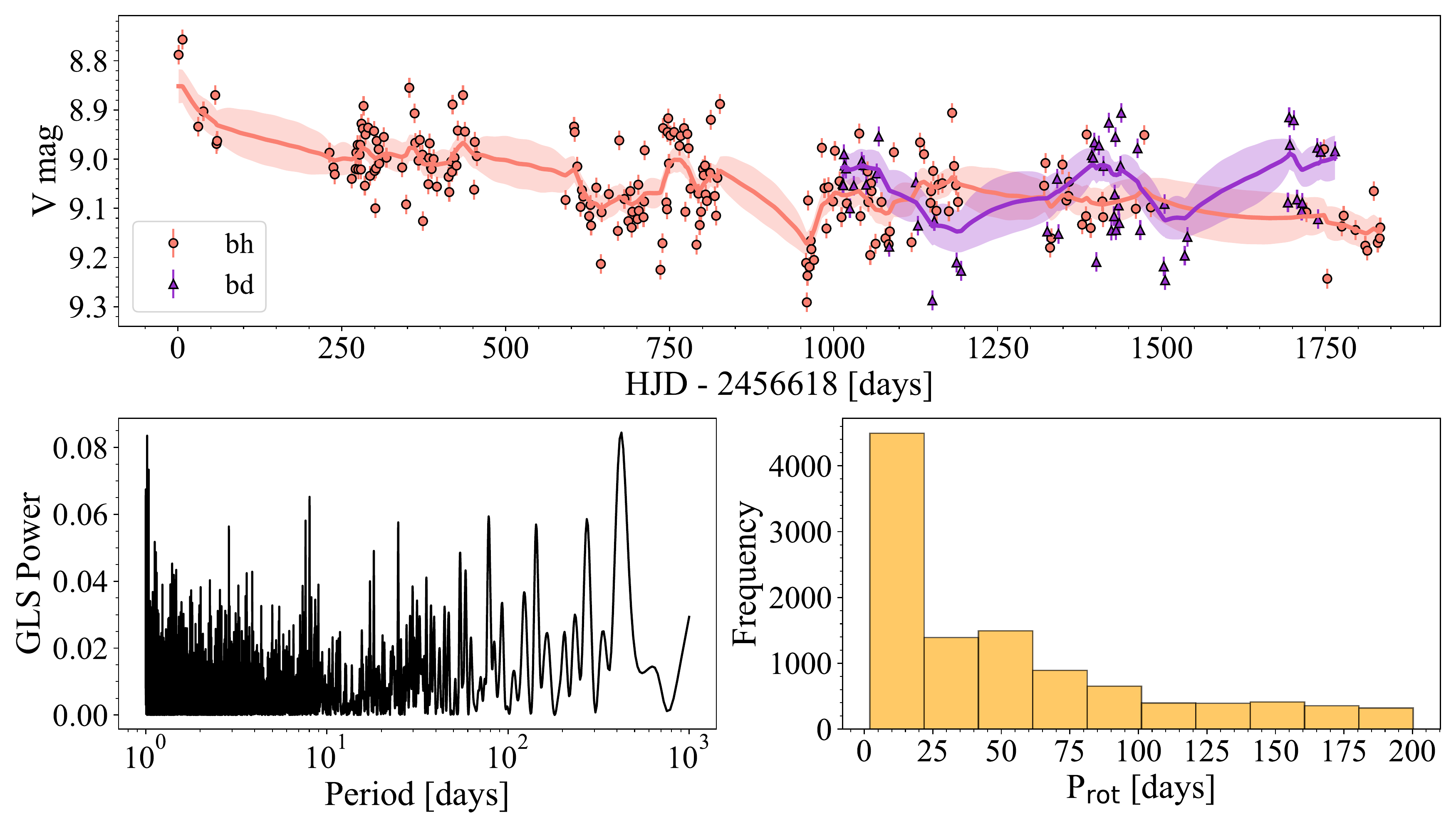}
   \caption{Ground-based long-term photometry of HD 20329. \textit{Top panel:} ASAS-SN $V$-band measurements from cameras \texttt{bd} and \texttt{bh}. \textit{Bottom left panel:} GLS periodogram (\citealp{Zechmeister2009}) of the $V$-band photometry. \textit{Bottom right panel:} Posterior distribution of the rotation period parameter from the fit. }
    \label{Fig:HD20329bASASSN}
\end{figure*}

Inhomogeneities in the stellar surface (e.g., spots or plages) that appear and disappear out of sight as the star rotates can imprint flux variations that can be detected by photometric measurements. We obtained $V$-band photometry for HD 20329b from ASAS-SN public light-curve archive\footnote{\url{https://asas-sn.osu.edu/}} (\citealp{Shappee2014}, \citealp{Kochanek2017}). The ASAS-SN archive has 251 observations spanning a time baseline of 1832 days ($t_{start} = 2456618.931$ HJD, $t_{end} = 2458451.844$ HJD). The ASAS-SN photometry is shown in Figure \ref{Fig:HD20329bASASSN}. The GLS periodogram of the photometry presents several peaks with about equal power. The strongest peak corresponds to a period of 423 days. We fit the ASAS-SN photometry using the package for Gaussian Processes \texttt{Celerite} (\citealp{ForemanMackey2017}) and fit the photometry with the kernel

\begin{equation}
    k_{ij\; \mathrm{Phot}} = \frac{B}{2+C} e^{-|t_i-t_j|/L} \left[  \cos \left( \frac{2 \pi |t_i-t_j|}{P_{rot}} \right) + (1 + C)  \right]
\label{Eq:StarRot_GPKernel}
,\end{equation}
where $|t_i-t_j|$ is the difference between two epochs of observations, $B$, $C$, $L$ are positive constants, and $P_{rot}$ is the stellar rotational period (see \citealp{ForemanMackey2017} for details). For the fit, we considered each ASAS-SN camera (\texttt{bd} and \texttt{bh}) as an independent instrument. Each data set had as free parameters the constants $B$, $C$, $L$ , but shared a common rotational period parameter.

The fitting procedure consisted of a global optimization of a log likelihood function. Then we sampled the posterior distribution of the kernel we used to fit the photometry with \texttt{Emcee} (\citealp{ForemanMackey2013}) using 120 chains and 10000 iterations for the main MCMC. The final parameter values (median and 1$\sigma$ uncertainties) were estimated from the posterior distribution.

Figure \ref{Fig:HD20329bASASSN} presents the photometry (top panel) and the posterior distribution of the fitted rotation period parameter (bottom right panel). The distribution of the rotation period parameter presents a long a tail of possible values, but the median value is $P_{rot} = 34.8^{+74.6}_{-30.8}$ days.

\section{Light curve and radial velocity joint fit}
\begin{figure*}[htbp]
   \centering
   \includegraphics[width=\hsize]{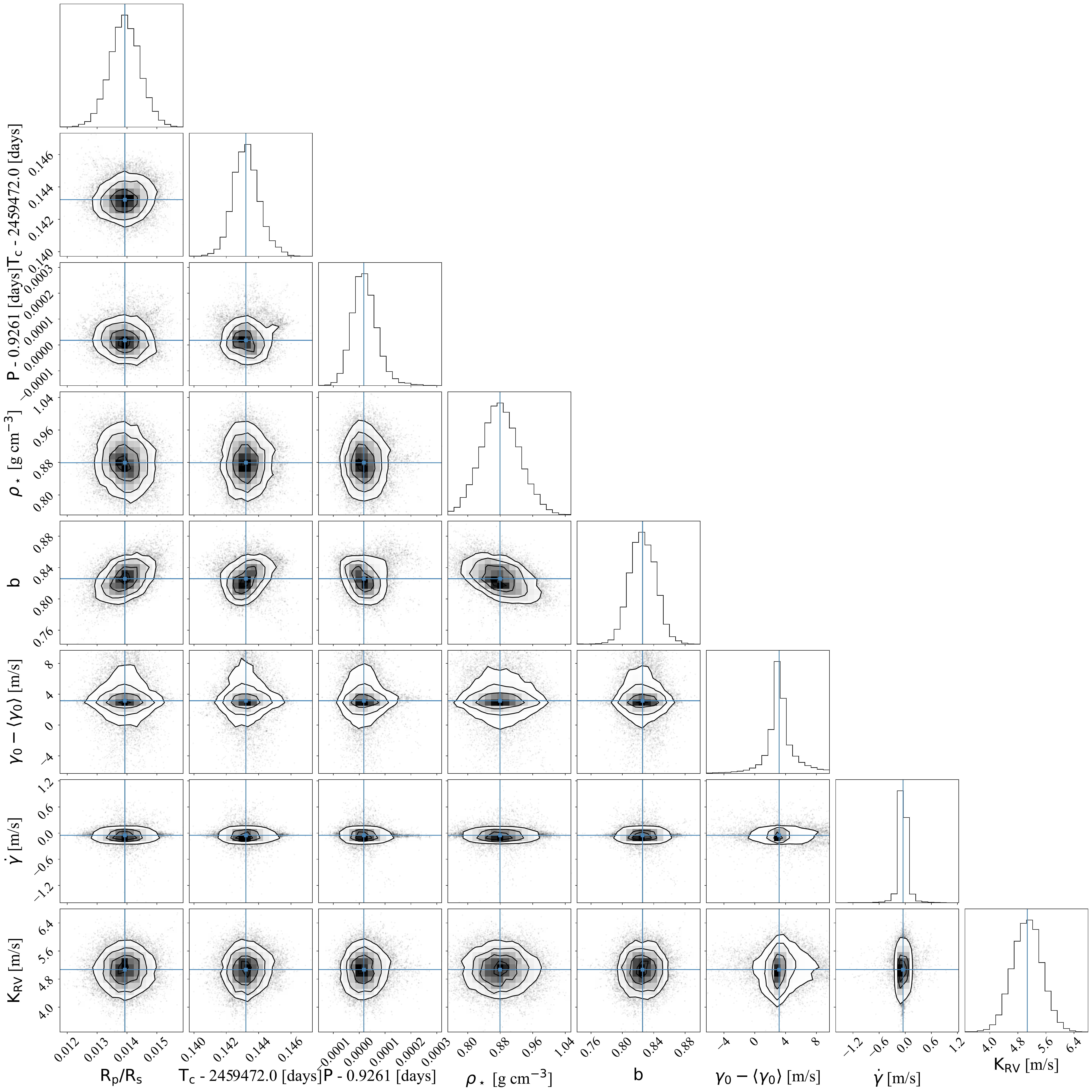}
   \caption{Correlation plot for the fit orbital parameters. Limb-darkening coefficients and systematic effect parameters were intentionally left out for easy viewing. The blue lines give the median values of the parameters.}
    \label{Fig:Fit_ParamDistr_CornerPlot}
\end{figure*}

In this section, we present the tests and results from the joint fit of \tess photometric data and the HARPS-N radial velocity measurements. Figure \ref{Fig:Fit_ParamDistr_CornerPlot} shows the parameter distributions for the best-fit model, table \ref{Table:Planet_parameters_Testmodels} presents the results of the fit models we tested in our model selection analysis, and table \ref{Table:Planet_parameters_curv} presents the results of a fit including a second-order term to model the linear trend seen in the RV residuals of the best-fit model.

\begin{table*}
\caption{HD 20329b fit parameters and final values for the other models tested in the joint fit.}
\label{Table:Planet_parameters_Testmodels}      
\centering                          
\begin{tabular}{l c c c}        
\hline\hline   
Parameter & $e=0$ (RVs without GPs) & $e \neq 0$ (RVs without GPs) & $e \neq 0$ (RVs with GPs) \\
\hline
\multicolumn{4}{c}{Fitted orbital and transit parameters} \\
\hline
$R_{p}/R_\star$ & $0.0139^{+0.0005}_{-0.0005}$ & $0.0139^{+0.0005}_{-0.0005}$ & $0.0140^{+0.0005}_{-0.0005}$  \\
$T_{c}$  [BJD] & $2459472.14344^{+0.00123}_{-0.00091}$ & $2459472.14401^{+0.00095}_{-0.00084}$ & $2459472.14327^{+0.00092}_{-0.00082}$ \\
$P$ [days] & $0.926218^{+0.000074}_{-0.000061}$ & $0.926116^{+0.000052}_{-0.000047}$ & $0.926118^{+0.000053}_{-0.000047}$ \\
$\rho_*$ [g cm$^{-3}$] & $0.88^{+0.05}_{-0.05}$ & $0.88^{+0.05}_{-0.05}$ & $0.88^{+0.05}_{-0.05}$ \\
$b$ & $0.837^{+0.022}_{-0.026}$ & $0.820^{+0.023}_{-0.025}$ & $0.831^{+0.020}_{-0.021}$ \\
$\sqrt{e}\cos(\omega)$ & --- & $0.283^{+0.054}_{-0.072}$ & $0.026^{+0.126}_{-0.127}$ \\
$\sqrt{e}\sin(\omega)$ & --- & $0.049^{+0.117}_{-0.135}$ & $-0.038^{+0.158}_{-0.149}$ \\
$\gamma_0 - \langle \gamma_0 \rangle$ [m/s] & $3.24^{+0.17}_{-0.18}$ & $3.39^{+0.19}_{-0.18}$ & $3.16^{+1.09}_{-1.02}$ \\
$\dot{\gamma}$ [m/s$^2$] & $-0.06^{+0.01}_{-0.01}$ & $-0.07^{+0.01}_{-0.01}$ & $-0.06^{+0.05}_{-0.05}$ \\
$K$ [m/s] & $4.99^{+0.25}_{-0.25}$ & $5.41^{+0.34}_{-0.33}$ & $5.08^{+0.42}_{-0.41}$ \\
$\sigma_{RV}$ [m/s] & $1.20^{+0.16}_{-0.14}$ & $1.18^{+0.15}_{-0.14}$ & $0.81^{+0.16}_{-0.14}$ \\
\hline
\multicolumn{4}{c}{Derived planet parameters} \\
\hline
R$_{p}$ [R$_{\oplus}$] & $1.73 \pm 0.07$ & $1.71 \pm 0.07$ & $1.73 \pm 0.07$ \\
M$_{p}$ [M$_{\oplus}$] & $7.42 \pm 1.13$ & $7.93 \pm 1.36$ & $7.43 \pm 1.27$ \\
$\rho_{p}$ [g cm$^{-3}$] & $7.81 \pm 1.57$ & $8.67 \pm 1.87$ & $7.94 \pm 1.68$ \\
$g_p$ [m s$^{-2}$] & $24.0 \pm 4.3$ & $26.5 \pm 5.1$ & $24.4 \pm 4.6$ \\
a [au] & $0.0180 \pm 0.0003$ & $0.0180 \pm 0.0003$ & $0.0180 \pm 0.0003$ \\
T$_{eq}$ ($A_B = 0.0$) [K] & $2140 \pm 27$ & $2139 \pm 28$ & $2140 \pm 28$ \\
T$_{eq}$ ($A_B = 0.3$) [K] & $1958 \pm 25$ & $1957 \pm 25$ & $1958 \pm 25$ \\
$\langle F_{p} \rangle$ [10$^5$ W/m$^2$] & $47.2 \pm 1.7$ & $47.2 \pm 1.7$ & $47.3 \pm 1.7$ \\
$S_{p}$ [$S_\oplus$] & $3470 \pm 124$ & $3465 \pm 124$ & $3472 \pm 124$ \\

\hline
\multicolumn{4}{c}{Fitted LD coefficients} \\
\hline
$q_{1\;TESS}$ & $0.32^{+0.02}_{-0.02}$ & $0.32^{+0.02}_{-0.02}$ & $0.31^{+0.02}_{-0.02}$ \\
$q_{2\;TESS}$ & $0.37^{+0.03}_{-0.03}$ & $0.37^{+0.03}_{-0.03}$ & $0.37^{+0.03}_{-0.03}$ \\
\hline
\multicolumn{4}{c}{Fitted GP parameters} \\
\hline
$\log(c_1)$ TESS S42 & $-7.95^{+0.08}_{-0.03}$ & $-7.95^{+0.09}_{-0.04}$ & $-7.97^{+0.07}_{-0.03}$  \\
$\log(\tau_1)$ TESS S42 & $0.36^{+0.25}_{-0.23}$ & $0.37^{+0.25}_{-0.24}$ & $0.34^{+0.25}_{-0.23}$  \\
$\log(c_1)$ TESS S43 & $-7.97^{+0.05}_{-0.02}$ & $-7.98^{+0.04}_{-0.02}$ & $-7.99^{+0.03}_{-0.01}$  \\
$\log(\tau_1)$ TESS S43 & $-0.32^{+0.13}_{-0.13}$ & $-0.34^{+0.13}_{-0.12}$ & $-0.35^{+0.14}_{-0.12}$  \\
$c_2$ & --- & --- & --- \\
$\tau_2$ & --- & --- & --- \\
\hline                                   
\end{tabular}
\tablefoot{The term $\dot{\gamma}$ was computed relative to $T_{\mathrm{base}} = 2459579.0$ BJD.}
\end{table*}

\begin{table*}
\caption{HD 20329b fit parameters (circular orbit), prior functions, and final values for the joint fit including a quadratic term in the RV measurements.}
\label{Table:Planet_parameters_curv}      
\centering                          
\begin{tabular}{l c c}        
\hline\hline   
Parameter & Prior & Value \\
\hline
\multicolumn{3}{c}{Fitted orbital and transit parameters} \\
\hline
$R_{p}/R_\star$ & $\mathcal{U}(0.005, 0.025)$ & $0.0140^{+0.0005}_{-0.0005}$ \\
$T_{c}$  [BJD] & $\mathcal{U}(2459471.7445,2459472.5445)$ & $2459472.14325^{+0.00081}_{-0.00077}$ \\
$P$ [days] & $\mathcal{U}(0.5,1.5)$ & $0.926108^{+0.000049}_{-0.000041}$ \\
$\rho_*$ [g cm$^{-3}$] & $\mathcal{N}(0.879,0.068)$ & $0.88^{+0.05}_{-0.05}$ \\
$b$ & $\mathcal{U}(0.0,1.0)$ & $0.826 \pm 0.017$ \\
$\gamma_0 - \langle \gamma_0 \rangle$ [m/s] & $\mathcal{U}(-6.30,9.70)$ & $3.70^{+3.20}_{-2.65}$ \\
$\dot{\gamma}$ [m/s$^2$] & $\mathcal{U}(-100.0, 100.0)$ & $-0.055^{+0.111}_{-0.115}$ \\
$\ddot{\gamma}$ [m/s$^3$] & $\mathcal{U}(-100.0, 100.0)$ & $-0.002^{+0.004}_{-0.006}$ \\
$K$ [m/s] & $\mathcal{U}(0.0,110.0)$ & $5.14 \pm 0.40$ \\
$\sigma_{RV}$ [m/s] & $\mathcal{U}(0.0,10.0)$ & $0.84^{+0.16}_{-0.15}$ \\
\hline
\multicolumn{3}{c}{Fitted LD coefficients} \\
\hline
$q_{1\;TESS}$ & $\mathcal{U}(0.0,1.0)$ & $0.32^{+0.02}_{-0.02}$ \\
$q_{2\;TESS}$ & $\mathcal{U}(0.0,1.0)$ & $0.37^{+0.02}_{-0.02}$ \\
\hline
\multicolumn{3}{c}{Fitted GP parameters} \\
\hline
$\log(c_1)$ TESS S42 & $\mathcal{U}(-8.0,2.3)$  & $-7.95^{+0.10}_{-0.04}$ \\
$\log(\tau_1)$ TESS S42 & $\mathcal{U}(-2.65,6.00)$ & $0.38^{+0.25}_{-0.24}$ \\
$\log(c_1)$ TESS S43 & $\mathcal{U}(-8.0,2.3)$ & $-7.98^{+0.04}_{-0.02}$ \\
$\log(\tau_1)$ TESS S43 & $\mathcal{U}(-2.65,6.00)$ & $-0.34^{+0.13}_{-0.13}$ \\
$c_2$ & $\mathcal{U}(0.0,100.0)$ & $4.85^{+7.30}_{-2.86}$ \\
$\tau_2$ & $\mathcal{U}(0.001, 150.0)$ & $5.16^{+7.20}_{-3.56}$ \\
\hline                                   
\end{tabular}
\tablefoot{$\mathcal{U}$, $\mathcal{N}$ represent uniform and normal prior functions, respectively. The terms $\dot{\gamma}$ and $\ddot{\gamma}$ were computed relative to $T_{\mathrm{base}} = 2459579.0$ BJD.}
\end{table*}

\section{Light-curve fit with TLCM}
\label{sec:appendix_tlcm}

We also fit the light curve with the {\sc {Transit and Light Curve Modeller}} code described in detail in \cite{csizmadia2020}. We briefly summarize its main features here. This code is able to perform a joint radial velocity and light-curve fit. The transit and occultation events can be modeled with different limb-darkening laws. The out-of-transit variations can be modeled by including the ellipsoidal, reflection, and beaming effects \citep[for details, see][]{csizmadia2021}. Contamination effects and eccentric orbit are included. The instrumental noise and stellar variability in the light curve are modeled by wavelets that are simultaneously fit with the light-curve model (for a more detailed description, see \citealp{csizmadia2021}).

In the fit, the free parameters were the scaled semi-major axis ($a/R_\mathrm{star}$), the planet-to-star radius ratio ($R_\mathrm{planet}/R_\mathrm{star}$), the impact parameter ($b$), the planet-to-star surface brightness ratio, the epoch, the period, the RV amplitude for the beaming effect, the planetary geometric albedo, the mass ratio for the ellipsoidal effect, the wavelet noise parameters $\sigma_w$ and $\sigma_r$, and the two limb-darkening coefficient combinations for the quadratic limb-darkening law $u_{+} = u_a + u_b$ and $u_{-} = u_a - u_b$. Since the derived eccentricity seems to be compatible with zero (see Table~\ref{Table:Planet_parameters}), we assumed a circular orbit for the planet.

The solution was optimized with the genetic algorithm \citep{geem} first, then a differential evolution MCMC analysis \citep{nelson14,sherri17} was performed to estimate the median of the posterior distribution. The uncertainties were obtained using the usual 16-84\% rule. The results are reported in Table~\ref{Table:Planet_parameters_Testmodels_TLCM}.

\begin{figure*}
   \centering
   \includegraphics[width=\hsize]{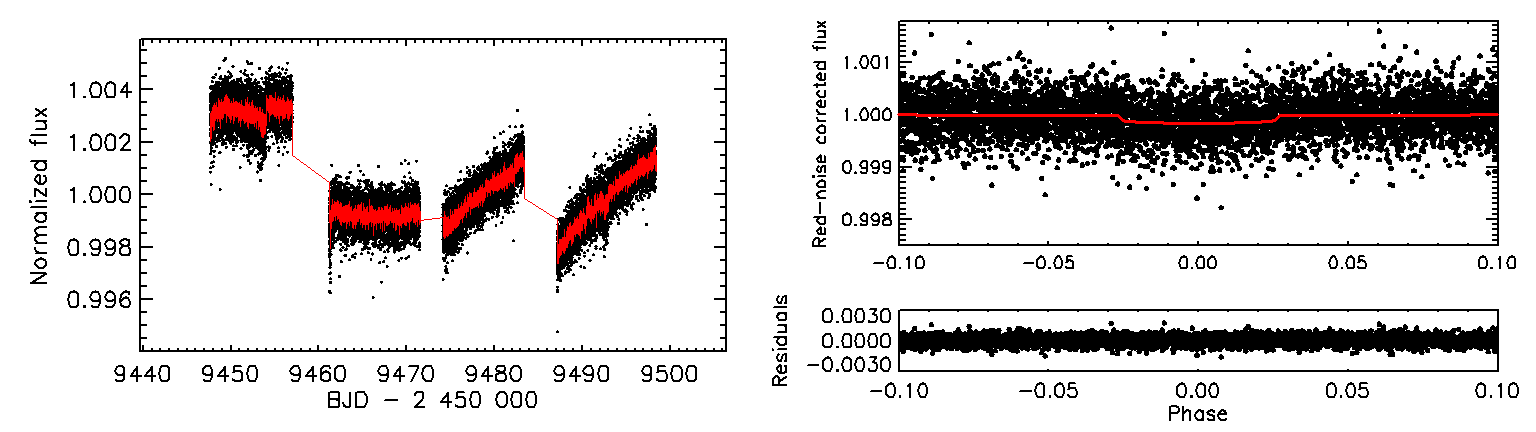}
   \caption{HD 20329 \textit{TESS} photometry and TLCM fit. \textit{Left panel:} \textit{TESS} normalized SAP-fluxes vs time. Black dots are the observations. The red curve is the result of the simultaneous transit + occultation + beaming + reflection + ellipsoidal + wavelet-based noise model fit. \textit{Top right panel:} Primary transit of HD 20329b (black dots) and best model fit (red line) after subtracting the red-noise component from the light curve (top). \textit{Bottom right panel}: Residuals of the fit.}
    \label{Fig:fig1_fig2_tlcm}
\end{figure*}

\begin{figure*}
   \sidecaption
   \includegraphics[width=\hsize]{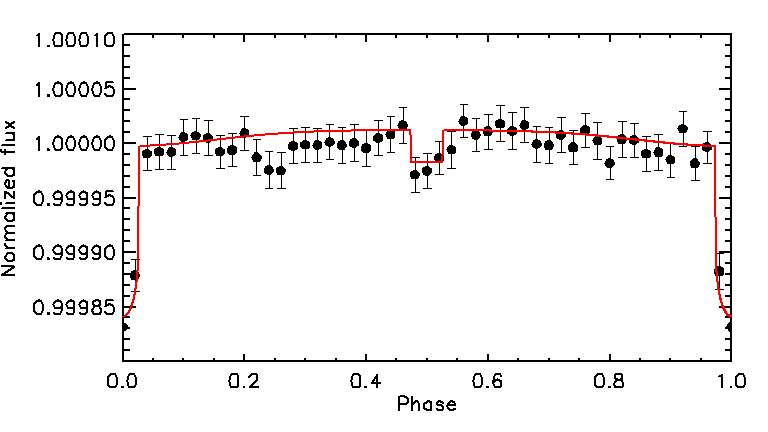}
   \caption{\textit{TESS} binned light curve of HD 20329b (50 bins with a $\sim$27 min. bin size), zoomed at the out-of-transit part (phase curve). See details in Appendix~\ref{sec:appendix_tlcm}. TLCM normalized the light curve to phase 0.25.}
    \label{Fig:fig4_tlcm}
\end{figure*}

The combined model + noise model fit is presented graphically in the left panel of Figure~\ref{Fig:fig1_fig2_tlcm}. For visualization purposes, we subtracted the wavelet-based red-noise component from the light curve, then we phase-folded and plotted it in the right panel of Figure~\ref{Fig:fig1_fig2_tlcm} . We binned the phase-folded, red-noise-corrected light curve into 50 bins (bin size $\sim$27 minutes) to show the occultation event (secondary transit) at phase 0.5. Although there is a flux loss at this phase, we did not detect the occultation of the planet at a $3\sigma$ significance level. This detection is just tentative. We argue for this in the following way: First, a similar flux loss is visible in this binned light curve at phase 0.25, where we do not expect such a flux drop (Figure~\ref{Fig:fig4_tlcm}). Second, the observed geometric albedo is $1.21^{+1.65}_{-1.71}$, meaning that it cannot be derived from this light-curve set. We also measured the surface brightness ratio of the planet and the star to be $I_\mathrm{planet}/I_\mathrm{star} = 0.09^{+0.08}_{-0.09}$ , which is compatible with zero. When blackbody radiation is asumed for the star and the planet, the median value would mean a temperature of 3300 K for the planet. This agrees well with the analysis presented in Section~\ref{sec:SecondaryTransit}. The occultation depth is $30 \pm 54$ ppm, which does not exclude a nondetection of the secondary transit.

We conclude that we can give only a $1\sigma$ upper limit of 84 ppm for the occultation depth in the system of HD 20329b with TLCM. The error bars given by TLCM are slightly larger than the values in Table~\ref{Table:Planet_parameters}. The derived planet-to-star radius ratio agrees within $1\sigma$ of the error bar with the values reported in Table~\ref{Table:Planet_parameters}, and all other parameters also agree reasonably well with the finally accepted values.

\begin{table*}
\caption{TLCM fit. $I_\mathrm{planet}/I_\mathrm{star}$ is the surface brightness ratio of the planet and the star, measured in the \textit{TESS} passband.}
\label{Table:Planet_parameters_Testmodels_TLCM}      
\centering                          
\begin{tabular}{l c c}        
\hline\hline   
Parameter & Prior & Circular orbit \\
\hline
\hline
$a/R_\mathrm{star}$                 & $\mathcal{U}(1,29)$ & $3.52^{+0.18}_{-0.17}$ \\
$R_\mathrm{planet}/R_\mathrm{star}$ & $\mathcal{U}(0,1)$  & $0.0128^{+0.0021}_{-0.0020}$ \\
$b$                                 & $\mathcal{U}(0,1)$  & $0.827^{+0.048}_{-0.067}$ \\
$I_\mathrm{planet}/I_\mathrm{star}$ & $\mathcal{U}(0,1)$  & $0.09^{+0.08}_{-0.09}$ \\
Epoch (BJD - 2 450 000)             & $\mathcal{U}(9472.141, 9472.145)$ & $9472.1429^{+0.0018}_{-0.0015}$ \\ 
Period (days)                       & $\mathcal{U}(0.925,0.927)$ & $0.926269^{+0.000183}_{-0.000250}$ \\       
$K_\mathrm{phot}$ (m/s)             & $\mathcal{U}(-1000,1000)$  & $19^{+39}_{24}$ \\
$M_\mathrm{planet}/M_\mathrm{star}$ & $\mathcal{U}(0.0, 0.02)$   & $0.00009^{+0.00013}_{_0.00008}$ \\
Geometric albedo of the planet      & $\mathcal{U}(-1,10)$      & $1.21^{+1.65}_{-1.71}$ \\
$\sigma_r$ (ppm)                    & $\mathcal{U}(0.0,10^6)$ & $27~465^{+416}_{-375}$ \\
$\sigma_w$ (ppm)                    & $\mathcal{U}(0.0,10^4)$ & $414^{+3}_{-3}$ \\
$u_{+}$                             & $\mathcal{U}(-1,2)$        & $0.55^{+0.75}_{0.95}$ \\
$u_{-}$                             & $\mathcal{U}(-1,2)$        & $0.05^{+0.99}_{0.76}$ \\
$\rho_{star}$ [g/cm$^3$]            & -              & $0.97\pm 0.12$ \\
Occultation depth (ppm)             & -              & $30\pm 54$ \\
\hline                                   
\end{tabular}
\end{table*}

\end{appendix}

\end{document}